\documentclass[10pt,journal]{IEEEtran}
\usepackage{amsmath,amssymb,graphicx,amsthm}
\usepackage{latexsym,todonotes}
\usepackage{mathtools}
\usepackage{cite}
\usepackage{color}
\usepackage{arydshln}
\usepackage{subfigure}
\usepackage{bm}
\usepackage[vlined,ruled]{algorithm2e}
\usepackage{setspace}
\SetAlCapSkip{0.5em}
\SetKwComment{Comment}{$\triangleright$\ }{}

\usepackage[bookmarks=false]{hyperref}

\usepackage[mathcal]{euscript}
\usepackage[deletedmarkup=xout]{changes}

\graphicspath{{./figs/}}

\newtheorem{theorem}{\bf Theorem}
\newtheorem{lemma}{\bf Lemma}
\newtheorem{proposition}[theorem]{\bf Proposition}
\newtheorem{corollary}{\bf Corollary}

\newtheorem{condition}{\bf Condition}

\makeatletter
\def\env@cases{%
  \let\@ifnextchar\new@ifnextchar
  \left.
  \def\arraystretch{1.2}%
  \array{@{}l@{\,\,}l@{}}%
}%
\makeatother

\begin{document}
\title{Energy-Efficient Job-Assignment Policy with Asymptotically Guaranteed Performance Deviation}
\author{Jing~Fu\IEEEauthorrefmark{1},~\IEEEmembership{Member,~IEEE},
Bill Moran\IEEEauthorrefmark{2},~\IEEEmembership{Member,~IEEE} \\
\IEEEauthorblockA{
\IEEEauthorrefmark{1}School of Mathematics and Statistics, the University of Melbourne, VIC3010, Australia\\
\IEEEauthorrefmark{2}Department of Electrical and Electronic Engineering, the University of Melbourne, VIC3010, Australia}\\
Email: jing.fu@unimelb.edu.au; wmoran@unimelb.edu.au
}

\maketitle 
\begin{abstract}%150~250 words
We study a job-assignment problem in a large-scale server farm system with geographically deployed servers as abstracted computer components (e.g., storage, network links, and processors) that are potentially diverse. We aim to maximize the energy efficiency of the entire system by effectively controlling carried load on networked servers. A scalable, near-optimal job-assignment policy is proposed. The optimality is gauged as, roughly speaking, energy cost per job. 
Our key result is an upper bound on the deviation between the proposed policy and the asymptotically optimal energy efficiency, when job sizes are exponentially distributed and blocking probabilities are positive. Relying on Whittle relaxation and the asymptotic optimality theorem of Weber and Weiss, this bound is shown to decrease exponentially as the number of servers and the arrival rates of jobs increase arbitrarily and in proportion.
In consequence, the proposed policy is asymptotically optimal and, more importantly, approaches asymptotic optimality quickly (exponentially). 
This suggests that the proposed policy is close to optimal even for relatively small systems (and indeed any larger systems), and this is consistent with the results of our simulations. 
Simulations indicate that the policy is effective, and robust to variations in job-size distributions.
\end{abstract}%193

%\bstctlcite{IEEEexample:BSTcontrol}

\IEEEpeerreviewmaketitle

\section{Introduction}\label{sec:introduction}

\IEEEPARstart{G}{lobal} Internet traffic is rapidly increasing, driving a parallel
growth in the data-center industry; over 500 thousand data centers
have been launched worldwide~\cite{state2011}. 
 According to an
estimate in 2013, about 91 billion kWh of electricity were consumed
by U.S.  data centers during that year, and the annual consumption
has been predicted to reach \$13 billion with nearly 100 million
metric tons of carbon pollution, potentially, by 2020~\cite{NRDC2014}. 
Servers are considered to be the major contributor to
the electrical consumption of data centers~\cite{kliazovich2015energy}.  We
study the dispatching policies for incoming jobs in a server
farm consisting of deployed servers/computer components, aiming to
maximize energy efficiency.

Energy-efficient scheduling policies for local server farms are
well-studied.
For instance, speed scaling technique can
reduce energy consumption by decreasing server speed(s)~\cite{
albers2014speed,lenhardt2017energy}; energy-efficient servers enable
dynamic right-sizing of server farms by powering servers on/off/into
power-conservative modes according to offered traffic load~\cite{lin2013dynamic,yao2014power}; and decision making
policies according to queue sizes of servers have been considered by
changing server working modes~\cite{
hyytia2014task, gebrehiwot2017near}. The development of distributed
cloud computing platforms has stimulated research on
geographically deployed energy-efficient server farms
\cite{lin2017game, shi2017online}.

Server farm vendors deploy a variety of computer components, such as
CPUs and disks, to meet various types of inquiries from Internet
users, and different generations of these components are present
simultaneously because of partial upgrading of old components and
purchasing new ones over time~\cite{guo2014server}.  
A diversity of physical computing/storage components are
available for use in cloud computing platforms, and are abstracted
(virtualized) as resources with varying attributes \cite{hameed2016survey}.  
All of these
have resulted in heterogeneity as an important feature in attempting
to undertake research on server farms,  
whereas~\cite{gandhi2011data,lu2013simple}
studied only identical servers.

Regardless of the complexity
caused by heterogeneity of server farms, 
large modern server farms with hundreds of thousands of computer components (abstracted servers) require all scheduling policies to be scalable.

Existing job-assignment policies have been discussed in server
farms that have negligible energy consumption on idle servers~\cite{albers2014speed,rosberg2014,hyytia2014task,fu2015energy}.
Power consumption on idle servers is normally significant in real
situations~\cite{barroso2007case}, and we so regard it in this paper.

On the other hand, job-assignment policies
for network resource allocation problems, such as~\cite{chowdhury2012vineyard,esposito2016distributed},
applicable for practical scenarios with heterogeneous servers
and jobs, were studied as static optimizations.
Profits  to be gained through dynamic release and reuse of  resources were ignored.
Here we use methods of stochastic optimization that capture dynamic properties of a system.

To maximize 
the energy efficiency, defined as the ratio of the long-run average throughput to the long-run average power consumption, in a stochastic system with heterogeneous servers, an asymptotically optimal
job-assignment policy was proposed in~\cite{fu2016asymptotic}.
This optimal policy approaches the optimal solution as the
numbers of servers in different server groups tend to infinity
proportionately. Nonetheless, the asymptotic optimality is
restricted in two aspects: a) although modern server farms are
normally large enough to be close to the asymptotic regime, the
critical value, above which the numbers of servers are
``sufficiently large'', remains unclear, and b) it was assumed that any
server in the server farm can serve any arriving job if it has a
vacant slot in its buffer. This constraint is not appropriate for geographically separated or functionally varied computer components (abstracted servers).
A detailed survey of other
published work is provided in Section~\ref{sec:rWork}.

We aim to maximize the energy efficiency and
study the deviation between a newly proposed, scalable
policy and the true optimal solution; particularly studying the
relationship between this deviation and the number of servers in the
system. Also, we extend the idea of~\cite{fu2016asymptotic} to a more general, realistic system 
in which the available servers for a given job are, to some extent,  job-dependent. 
This extension significantly complicates the entire system and enables 
consideration of the effects on the problem of the geographical locations and service features (e.g., CPU or memory) of computer components. This was not captured in \cite{fu2016asymptotic}.
We refer to the abstracted servers that are potentially able to serve a job as the \emph{available servers} of this job.

The primary contribution of this paper is a sharpening of the asymptotic optimality results in a heterogeneous server farm, discussed in a special case in \cite{fu2016asymptotic}.  Specifically, we prove that, when the job
sizes are exponentially distributed and the blocking probabilities of jobs are always positive, there is a hard upper bound on
the deviation between a simple, scalable policy and the optimized
energy efficiency in the asymptotic regime; this upper bound \emph{diminishes
exponentially} as the number of servers in server groups and the
arrival rates of jobs tend to infinity proportionately. 
In other words, the scalable policy approaches asymptotic optimality quickly (exponentially) as the size of the optimization problem increases.
We refer to this upper bound as the \emph{deviation bound}, and the policy as  \emph{Priorities accounting for Available Servers} (PAS), as it is a priority-style policy and applicable for a system with different sets of available servers.

Our secondary contributions are twofold:
\begin{itemize}
\item 
We consider a large-scale system, potentially containing several geographically distributed server farms, and regard this hybrid system as an abstracted server farm model.
We propose a scalable PAS policy in this server farm with heterogeneous servers and jobs, where energy consumption and service rates of servers can be arbitrarily related.  
As mentioned in our primary contribution, the PAS policy is proved to rapidly (exponentially) approach asymptotic optimality as the scale of the server farm increases.
To the best of our knowledge, no existing work has studied the energy efficiency of such a realistically scaled heterogeneous server farm, nor does any existing work propose a scalable scheduling policy with proven guaranteed performance.

\item By numerical simulations, we demonstrate that PAS is nearly optimal even for relatively small server farms.
Together with the rapidly decreasing deviation bound, when job sizes are exponentially distributed and the blocking probabilities of jobs are always positive, it is likely to be near-optimal in all larger server farms and proved to approach optimality as the server farm sizes tend to infinity.
In particular, the deviation bound of PAS is
    demonstrated to decrease with increasing server farm size,
    consistent with our theoretical results, and to be less than $3\%$
    in all our experiments involving only 100 servers (computer components).  Also, we
    numerically demonstrate that the PAS policy is relatively
    insensitive to the specific job-size distribution by comparing its
    energy efficiency with different job-size distributions.
\end{itemize}

The paper is organized as follows:
Section~\ref{sec:rWork}: discussion of related work on job assignment
policies; Section~\ref{sec:model}: description of the server farm
model; Section~\ref{sec:problem}: definition of the stochastic
optimization problem; Section~\ref{sec:index}: description of the
PAS policy; Section~\ref{sec:deviation}: proof of the
deviation bound; Section~\ref{sec:numerical}: numerical results;
Section~\ref{sec:conclusion}: conclusions.

\vspace{-0.3cm}
\section{Other Related Work}\label{sec:rWork}

Asymptotically optimal job-assignment policies applicable to a parallel queueing model with infinite buffer size on each
server were studied in
\cite{ayesta2013scheduling,verloop2015asymptotically}.  
In~\cite{gupta2011stochastic,sakuma2010asymptotic},
policies were proposed for server farms without capacity constraints
(with infinite buffer size), aimed at minimization of average delay.
A detailed survey
of asymptotically optimal job-assignment policies was given in~\cite{fu2016asymptotic}.

In particular, the asymptotic results presented in this paper are obtained by implementing the Whittle relaxation technique, which was originally designated for the \emph{restless multi-armed bandit problem} (RMABP) \cite{whittle1988restless}.
The optimization of a general RMABP was proved to be PSPACE-hard \cite{papadimitriou1999complexity}; nonetheless, in \cite{weber1990index}, a scalable policy, proposed through the Whittle relaxation technique  and referred to as the \emph{(Whittle) index policy} in \cite{whittle1988restless}, was proved to be asymptotically optimal under conditions that require a global attractor of a stochastic process associated with the RMABP, and \emph{Whittle indexability} \cite{whittle1988restless}.
There is not a necessary implication between one of these conditions and asymptotic optimality.
In \cite{nino2001restless,nino2002dynamic}, Ni\~no-Mora proposed and analyzed the \emph{Partial Conservation Law (PCL) indexability} for the performance of scheduling problems, such as the (restless) multi-armed bandit problem. This implies (and is stronger than) the Whittle indexability. A More detailed survey about indexability can be found in \cite{nino2007dynamic}.

Other publications that discuss management of jobs in server farms by
distributing offered traffic were published
recently~\cite{lenhardt2017energy,lin2017game}.  Lenhart \emph{et
  al.}~\cite{lenhardt2017energy} provided an experimental analysis on
energy-efficient web servers with an assumption of a cubic
relationionship between server power consumption and traffic load. The
number of servers (nodes) tested in~\cite{lenhardt2017energy} is very
small compared to a real system in modern data centers.  Lin \emph{et
  al.}~\cite{lin2017game} analyzed energy efficiency performance
in communication networks in a server farm model. They assumed power
consumption is  negligible on idle servers (multiplexing/aggregating
nodes in network).
Recall that, here, we allow the possibility of positive power consumption on idle.

An auction based mechanism was studied in a server farm model with
heterogeneous servers (resources) and job types (users) in~\cite{shi2017online}.  The authors provided a worst-case
ratio (competitive ratio) of the revenue (social welfare) under their
proposed policies relative to the optimal solution, and
showed it to be linearly increasing in the time horizon. Also, in~\cite{shi2017online}, the length of each job is assumed  known
before assigning it.  Here, we consider the more realistic situation
that the job length remains unknown until it is completed.

Stochastic job-assignment techniques were studied in~\cite{wei2017data}, where  a linearly increasing
relationship of  reward/cost rates to  traffic loads of servers is assumed.
They proposed a reasonable and scalable policy with a given parameter $V$ and proved the deviation between this policy and the optimal solution to be $O(1/V)$, where $V$ is a parameter related to the Lyapunov optimization technique. 
For the version of energy-efficient server farms considered here,
the energy consumption rate of computer components is generally non-linear in its traffic load, so the linearity is not appropriate for modeling 
energy consumption rates of computer components in Cloud environments.
In consequence, the results in~\cite{wei2017data} are not directly
applicable to the energy-efficient server farm.

In~\cite{wei2017data},  the blocking of jobs was allowed when not all servers are fully occupied.
Also, a deterministic job lifespan model was assumed and all server were assumed to be able to handle all jobs. We strictly prohibit  blocking of jobs when available servers are not fully occupied, to ensure the fairness for all customers. We also consider a diversity of jobs with randomly generated job sizes (remaining unknown until completed) and with job-dependent sets of available servers.

In summary, the non-linearity of power functions and the complexity of
our server farm model prevent applicability of existing methods  from being direct.  Also, there is
no published work that provided theoretical
bounds that are quickly (exponentially) decreasing in the number of servers
between proposed policy and the optimal solution.

Moreover, as mentioned in Section~\ref{sec:introduction}, powering servers on/off \cite{gandhi2011data,lin2013dynamic,lu2013simple} or switching into power conservative modes with additional suspending time \cite{yao2014power,hyytia2014task,wei2017data,lin2017game} 
enables dynamic variation of the size/number of working of working servers in a server farm. In \cite{lin2013dynamic,lu2013simple}, such server farms are called \emph{dynamic right-sizing} server farms. In similar vein to \cite{rosberg2014,fu2015energy,fu2016asymptotic}, for the purposes of this paper, a fixed number of working servers is postulated in a server farm with no possibility of powering off or state switching, with substantial delay, during the time period under consideration.
In practice, this corresponds to periods during which no powering off or state switching, with concomitant substantial delays, takes place. In this context, the job assignment policies discussed here can be combined with the right-sizing techniques, as appropriate. Note  that frequent powering off/on or state switches increases wear and tear of hardware 
and ensuing requirements for costly replacement and maintenance \cite{pore2015techniques}.

\vspace{-0.3cm}
\section{Model}\label{sec:model}

We classify incoming jobs into $J$ job types labeled by an integer $j$
($j=1,\ldots,J$), each of which has an arrival rate $\lambda_j>0$ indicating the
average number of arrivals per unit time, following a Poisson
process as previous studies in Cloud
environments
\cite{li2017quantitative,sebastio2018holistic}.
If  groups/type of Internet/network customers decide to send
requests independently and
identically during a given time period and the number of such customers is
sufficiently large for the  corresponding dynamic process to become
stationary, then it is reasonable to model the arrival process of
customer requests for this type as a Poisson process, although
the arrival rates may vary from one time period to another, which is consistent with observations of real-world tracelogs~\cite{guo2018optimal,reiss2012heterogeneity}.

These jobs will be undertaken by servers or blocked.
We classify the servers into different server groups according to their functional features and  profiles. Define the set of server groups as $\mathcal{K}\coloneqq \{1,2,\ldots,K\}$.
We assume that there are in total $K\ge 2$ server groups and $R_k\geq 1$ servers
in group $k$. 
Each job type $j$ is 
only able to be served by a server from a subset
$\emptyset\ne \mathcal{K}_j\subset \mathcal{K}$ of server groups. We
say that $k\in \mathcal{K}_j$ is an \emph{available server group } for
job type $j$ and a server of group $k$ is an available server for a job of type $j$.
A server's availability of serving different jobs can be affected by its functional features,  jobs' preferences, and geographical distances from the jobs.

The \emph{sizes} of jobs of the same type are independent identically
distributed (i.i.d.) with average job size normalized to be
one (bit). 
We assume, for convenience, i.i.d. job sizes across all job types for the theoretical development, but we provide numerical results in Section~\ref{sec:numerical} to indicate the robustness of our results when when job-size distributions vary across job types.

A server of group $k$ serves its jobs at a total and peak rate of
$\mu_k$ using the processor sharing (PS) service discipline.
When the server is not idle, the service rate received by each job  is then $\mu_k$ divided by the number of jobs in the server buffer.
The service rate of each server in group $k$ is supported by consuming non-negligible amount of energy, even in its idle mode.
Similarly, we assume that each server operates in a power-consuming mode with its peak energy consumption rate when there are jobs accommodated. We refer to this peak energy consumption rate as the energy consumption rate of a busy server, denoted by $\varepsilon_k$ for group $k=1,2,\ldots,K$.
When the server is idle, it automatically changes to a power-conservative mode and consumes $\varepsilon_k^0$ energy per unit time. 
Evidently, $\varepsilon_k>\varepsilon_k^0\geq 0$.
The service rates and energy consumption rates of busy/idle servers are intrinsic parameters
determined  by the server hardware features and profiles, and
the relationship between them can be arbitrary in this paper.

As in \cite{lin2013dynamic,hyytia2014task,rosberg2014,fu2015energy,fu2016asymptotic},  the busy/idle operating rule is more appropriate to machines working in two-power modes, such as Oracle Sun Fire X2270 M2, Cisco UCS C210, Cisco MXE 3500 and Cisco UCS 5108.
Also, the IEEE 802.11 standards define exactly two power saving modes for network components in energy-conservative communications.

Moreover, we study a system with finite service capacity on each server, referred to as its \emph{buffer size}. 
It provides a finite bound on the number of jobs being simultaneously served by this server. 
Let $B_k\geq 1$ represent the buffer size of a server in group $k$.

A \emph{policy} $\phi$ is a mechanism to assign an arriving job of
type $j$ to a server of group $k\in \mathcal{K}_j$ with at least one
vacant slot in its buffer. 
A fairness criterion requires equal treatment of the different job
types.  In particular, rejection of jobs sent by different
users is
not permitted if there are vacant slots on available servers.  If
there is no such server (all available  buffers are fully
occupied), the arriving job is lost.

Consider a
ratio of total arrival rate to the total service rate, i.e.,
$\rho \coloneqq
\sum_{j\in\mathcal{J}}\lambda_{j}/\sum_{k\in\mathcal{K}}\mu_{k}$, the
\emph{normalized offered traffic} (see~\cite{fu2016asymptotic}).
  For
a specific job type $j$, the \emph{normalized offered traffic of type
  $j$},
$\rho_{j}\coloneqq \lambda_{j}/\sum_{k\in\mathcal{K}_{j}}\mu_{k}$.

We assume the existence of long-run averages of throughput and power
consumption, and refer to them as the \emph{job throughput} and the
\emph{power consumption} of the system,
respectively. Precise definitions are given in Section~\ref{sec:problem}. 
We define $\mathcal{L}^{\phi}$ and
$\mathcal{E}^{\phi}$ to be  the job throughput and power consumption of
the system under policy $\phi$, respectively. The \emph{energy efficiency of policy $\phi$},
$\mathcal{L}^{\phi}/\mathcal{E}^{\phi}$, is  the
objective of our problem for energy-efficient server farms.
The value of the energy efficiency indicates the average job throughput achieved by consuming one unit energy. Since we do not permit rejection of jobs when there are vacancies on available servers, the objective encapsulates a trade-off between performance and energy consumption.

\vspace{-0.3cm}
\section{Stochastic Optimization Problem}\label{sec:problem}
We study a stochastic system, where the dynamically released service capacities of physical components can be reused. 
To capture the stochastic features, we start by introducing the stochastic state of the entire system: a server farm with tens of thousands of heterogeneous servers.

Define the \emph{state of an individual server} as the number of jobs currently on the
server.  The set of all possible states of a server of group
$k\in\mathcal{K}$, is denoted by
$\mathcal{B}_{k}=\{0,1,\ldots,B_{k}\}$, where $B_{k}\geq 1$.  As in
\cite{fu2016asymptotic}, states $0,1,\ldots,B_{k}-1$ are called
\emph{controllable}, and the state $B_{k}$ is \emph{uncontrollable},
since all new jobs will be rejected by a server in state $B_{k}$.
Denote the set of all controllable states of server group $k$ by
$\mathcal{C}_k=\{0,1,\ldots,B_{k}-1\}$ and that of the
uncontrollable states by $\mathcal{U}_k=\{B_{k}\}$.

Servers of the same group have potentially different states
in the stochastic process.  Define the set of servers
in group $k$ as 
$\mathcal{R}_k$.  The set of all
servers is
$\mathcal{S} \coloneqq \bigcup_{k\in\mathcal{K}}\mathcal{R}_{k}$, 
the
set of servers available for jobs of type $j\in\mathcal{J}$ is
$\mathcal{S}_{j}\coloneqq
\bigcup_{k\in\mathcal{K}_{j}}\mathcal{R}_{k}$, 
and
the state space of the entire system is
$\mathcal{B}\coloneqq
\prod_{k\in\mathcal{K}}(\mathcal{B}_{k})^{R_{k}}$. 
The size of the
state space thus increases exponentially in the number of servers in
the server farm, i.e., $|\mathcal{S}|$, which, in itself, is normally
very large in modern real server farms.

Decisions driven by  a stationary policy $\phi$ applied to job arrivals
rely on the state of the system just before an arrival
occurs and the information known in this state, such as average rates of transitioning to other states.
For a policy $\phi$, $s\in\mathcal{S}$,
$j\in\mathcal{J}$, we define the action variable
$a^{\phi}_{j,s}(\bm{n})\in\{0,1\}$, $\bm{n}\in\mathcal{B}$, to indicate
the decision under policy $\phi$ for an arriving job of type $j$ on
server $s$ in state $\bm{n}$: server $s$ accepts an
arriving job of type $j$ if $a^{\phi}_{j,s}(\bm{n}) = 1$; and does not
accept any job otherwise.
In this context, for each job of type $j$, there is at most one server $s$ with $a^{\phi}_{j,s}(\bm{n}) = 1$ among all available servers.

In addition, we define, for $\bm{n}\in\mathcal{B}$, $j\in\mathcal{J}$, $s\in\mathcal{S}$,
\begin{itemize}
\item
$a^{\phi}_{j,s}(\bm{n})\equiv 0$ if $s\notin\mathcal{S}_{j}$ to prevent a server that is unavailable for job type $j$;
\item $a^{\phi}_{j,s}(\bm{n})\equiv 0$ if $n_{s}=B_{k}$,
  $s\in\mathcal{R}_{k}$, $k\in\mathcal{K}_{j}$, to prevent a server
  from accepting a new job when it is fully occupied. 
\end{itemize}
Then, the action space, as the discrete set of possible values of the action variables, is
\begin{equation}\label{eqn:action_space}
\mathcal{A} \coloneqq \prod_{j\in\mathcal{J}}\{0,1\}^{\sum_{k\in\mathcal{K}_j}R_k},\vspace{-0.2cm}
\end{equation}
where we recall that $\mathcal{K}_j$ is the set of available server groups for job type $j$ and $R_k$ is the number of servers in group $k$.
With large number of servers $R_k$, multiple job types (i.e., $J>1$) significantly enlarge the number of action variables in parallel with the size of the action space.

Let $N_{s}^{\phi}(t)\in \mathcal{B}_{k}$, $s\in\mathcal{R}_{k}$,
$k\in\mathcal{K}$, represent the state of server $s$ at time $t$ under
policy $\phi$, and
$\bm{N}^{\phi}(t)=(N^{\phi}_{s}(t):s\in\mathcal{S}) \in \mathcal{B}$. 
For simplicity, we always consider an empty system at time $0$, $\bm{N}^{\phi}(0)=\bm{0}$.

It will be convenient to consider mappings
$\bm{f}=(f_{1},f_{2},\ldots,f_{K})$, where
$f_k:\mathcal{B}_k\rightarrow \mathbb{R}$,
We refer to
such a vector of mappings $\bm{f}\in\prod_{k\in\mathcal{K}}\mathbb{R}^{\mathcal{B}_k}$ as
the vector of \emph{reward rate functions}.  Then, for a given
$\bm{f}\in \prod_{k\in\mathcal{K}}\mathbb{R}^{\mathcal{B}_k}$, for
some $k$, we define the \emph{long-run average performance} of the
system under policy $\phi$ to be
\begin{equation}\label{definition:long-run-reward}
\Gamma^{\phi}(\bm{f})=\lim\limits_{t\rightarrow +\infty}\frac{1}{t}\mathbb{E}\left[\int_{0}^{t}\sum\limits_{k\in\mathcal{K}}\sum\limits_{s\in\mathcal{R}_{k}}f_{k}(N^{\phi}_{s}(u))du\right],
\end{equation}
where we assume the existence of such limit.

Specifically, we define $f^{\mu}_{k}(n_{k})$ and
$f^{\varepsilon}_{k}(n_{k})$, $n_{k}\in\mathcal{B}_{k}$,
$k\in\mathcal{K}$, as the service rate and energy consumption rate of
a server in group $k$ in state $n_{k}$, respectively, and consider
them as the reward rate functions; that is, for $k\in\mathcal{K}$, 
$f^{\mu}_{k}$ and $f^{\varepsilon}_{k}$ are mappings: $\mathcal{B}_k\rightarrow\mathbb{R}$.
As
defined in Section \ref{sec:model}, $f^{\mu}_{k}(n_{k}) = \mu_{k}$,
$f^{\varepsilon}_{k}(n_{k})=\varepsilon_{k}$ for $n_{k}>0$,
$f^{\mu}_{k}(0)=0$ and $f^{\varepsilon}_{k}(0)=\varepsilon_{k}^{0}$,
where $\mu_{k}>0$, $\varepsilon_{k}>\varepsilon_{k}^{0}\geq 0$,
$k\in\mathcal{K}$.  For the vectors
$\bm{f}^{\mu}=(f^{\mu}_{1},f^{\mu}_{2},\ldots,f^{\mu}_{K})$ and
$\bm{f}^{\varepsilon}=(f^{\varepsilon}_{1},f^{\varepsilon}_{2},\ldots,f^{\varepsilon}_{K})$,
the job throughput of the entire system is, then,
$\Gamma^{\phi}(\bm{f}^{\mu})$, and the power consumption of the system
is $\Gamma^{\phi}(\bm{f}^{\varepsilon})$.
Recall that our objective is to maximize the energy efficiency of the entire system; that is, $\mathcal{L}^{\phi}/\mathcal{E}^{\phi}=\Gamma^{\phi}(\bm{f}^{\mu})/\Gamma^{\phi}(\bm{f}^{\varepsilon})$.

To complete necessary constraints on the action variables of our optimization problem, there remain definitions to accommodate  behavior that \emph{blocks} an arriving job.
We define a virtual server group $0$ with server number
$R_{0}=1$ and server set $\mathcal{R}_{0}=\{1\}$, which
receives blocked jobs.  Any server of group $0$ has only one state
with zero transition rate all the time: that is, it does not generate
any reward or cost to the entire system.  We define $\mathcal{B}_{0}$ as the state space of a server of group
$0$ where $|\mathcal{B}_{0}|=1$. Also, the set of controllable
states for $0$ is set to be
$\mathcal{C}_{0}=\mathcal{B}_{0}$ and the one for
uncontrollable states is $\mathcal{U}_{0} = \emptyset$.

We extend the original definition of a policy $\phi$ determined by
actions $a^{\phi}_{j,s}(\bm{n})$  ($\bm{n}\in\mathcal{B},\
s\in\mathcal{S}, \ j\in\mathcal{J}$), to that
determined by actions $(a^{\phi}_{j,s}(\bm{n})$:\
$\bm{n}\in\mathcal{B},\ s\in\mathcal{S},\
j\in\mathcal{J};\ a^{\phi}_j(\bm{n}))$, where $a^{\phi}_j(\bm{n})\in\{0,1\}$ represents the action
variable for the only server of virtual group $0$ for job type $j$.  We
slightly abuse the notation and still use $\phi$ to denote
such a policy.

Let $k(s)$ be the label of the server group satisfying
$s\in\mathcal{R}_{k(s)}$,
and $I(x)$ be the Heaviside function: for $x\in\mathbb{R}$, 
$I(x) =1$ if $x>0$; and $0$ otherwise.
Define $\Theta(x) = xI(x)$, $x\in\mathbb{R}$.  
Our problem  is then encapsulated by
\vspace{-0.2cm}
\begin{equation}
\max_{\phi} \Gamma^{\phi}(\bm{f}^{\mu})/\Gamma^{\phi}(\bm{f}^{\varepsilon})\label{eqn:objective}
\vspace{-0.2cm}
\end{equation}
with policy $\phi$ subject to
\begin{equation}
\sum\limits_{s\in\mathcal{S}_{j}}a^{\phi}_{j,s}\bigl(\bm{N}^{\phi}(t)\bigr) 
+ \Theta\Bigl(\overline{a}^{\phi}_{j}\bigl(\bm{N}^{\phi}(t)\bigr) \Bigr)
= 1, 
\forall j\in\mathcal{J},
t\geq0,
\label{eqn:org:constraint_1}
\end{equation}
\begin{equation}
\overline{a}^{\phi}_{j}(\bm{N}^{\phi}(t))+ \sum\limits_{s\in\mathcal{S}_{j}}I\left(B_{k(s)}-N^{\phi}_{s}(t)\right) \leq 1 ,
\forall 
j\in\mathcal{J},
t\geq 0,
\label{eqn:org:constraint_2}
\vspace{-0.3cm}
\end{equation} 
by introducing variables $\overline{a}^{\phi}_{j}(\bm{n})\in\mathbb{R}$ and setting $a^{\phi}_{j}(\bm{n}) = \Theta(\overline{a}^{\phi}_{j}(\bm{n}))$, $\bm{n}\in\mathcal{B}$. 
Define $\Phi$ to be the set of all the policies $\phi$ satisfying \eqref{eqn:org:constraint_1} and \eqref{eqn:org:constraint_2}.
Constraints in \eqref{eqn:org:constraint_2} and variables $\overline{a}^{\phi}_{j}(\bm{n})$ are introduced to guarantee that jobs of type $j$ are blocked if and only if all the available servers are fully occupied:
\begin{itemize}
\item if all available servers for job type $j$ are fully occupied at time $t$, then, because all available servers are in uncontrollable states, 
$a^{\phi}_{j,s}(\bm{N}^{\phi}(t))=0$ for all $s\in\mathcal{S}_j$ and 
the constraint in \eqref{eqn:org:constraint_1} forces $a^{\phi}_j(\bm{N}^{\phi}(t)) =\Theta(\overline{a}^{\phi}_{j}(\bm{N}^{\phi}(t)))= 1$, which does not violate the constraint in \eqref{eqn:org:constraint_2}; 
\item otherwise, the constraint in \eqref{eqn:org:constraint_2} forces $\overline{a}^{\phi}_{j}(\bm{N}^{\phi}(t))\leq 0$, which leads to $a^{\phi}_j(\bm{N}^{\phi}(t)) = \Theta(\overline{a}^{\phi}_{j}(\bm{N}^{\phi}(t))) = 0$ and so a newly-arrived job of type $j$ cannot be blocked.
\end{itemize}

We aim at a largely scaled problem that exhibits
inevitably high computational complexity.
A special case of our problem is in fact an instance of the Restless Multi-Armed Bandit Problem (RMABP)~\cite{whittle1988restless}.
The RMABP has been proved to be PSPACE-hard~\cite{papadimitriou1999complexity} in the general case, so that near-optimal, scalable approximations
are the impetus to this paper.
Here, we consider a general case involving not only heterogeneous servers, but also heterogeneous jobs, which is much more practical and  significantly increases the size of the action space, as described in \eqref{eqn:action_space}.
In particular, different sets of available servers for different jobs enable consideration of geographically distributed servers, jobs' preferences and functional differences among servers.

\vspace{-0.3cm}
\section{Priority-Style Policy}\label{sec:index}

\vspace{-0.1cm}

As mentioned in Section~\ref{sec:introduction},
in \cite{fu2016asymptotic}, a policy that always prioritizes the most efficient servers was proposed and proved to approach the optimality when the problem size (number of servers) becomes arbitrarily large; in other words, it is asymptotically optimal.
However, this policy requires that the set of available servers for each job always include all servers in the server farm, and the asymptotic optimality is not applicable in a realistic system with a large but finite number of servers. It is important to know the detailed relationship between the performance degradation and the problem size.

Recall our objective of maximizing the energy efficiency of the entire server farm, defined as the ratio of the job throughput to the power consumption. 
The power consumption can be interpreted as the cost used to support corresponding job throughput.
In this context, for an idle server in group $k\in\mathcal{K}$, $\varepsilon_k^0$ units of power are consumed in support of a zero service rate; if the server becomes busy,  $\varepsilon_k-\varepsilon_k^0$ power is added to produce a service rate $\mu_k$ .

In other words, the idle power $\varepsilon_k^0$ is a persistent and uncontrollable cost producing no service rate; while, $\varepsilon_k-\varepsilon_k^0$ is the productive and controllable part of the power that serves jobs at service rate $\mu_k$. 
We propose a policy that always
prioritizes servers producing higher service rates per unit controllable power; namely, the ratio of its service rate to the productive part of its power consumption, $\mu_{k}/(\varepsilon_{k}-\varepsilon_k^0)$. The ratio was referred to as the  \emph{effective energy efficiency} in \cite{fu2016asymptotic}.

In particular, for an incoming job of type $j\in\mathcal{J}$ with a set of available servers $\mathcal{S}_j$, the job will be assigned to a server in $\mathcal{S}_j$ with highest effective energy efficiency and with at least one vacancy in its buffer. As indicated earlier, we refer to such a policy as the \emph{Priorities accounting for Available Servers} (PAS).
Note that PAS uses a similar idea to that proposed in \cite{fu2016asymptotic}, but is applicable for our problem with different sets of available servers.

\IncMargin{-0.5em}

\begin{algorithm}[t]
\linespread{0.6}\selectfont

\SetKwFunction{FPAS}{UpdatingUponArrival}
\SetKwProg{Fn}{Function}{:}{\KwRet}
\SetKwInOut{Input}{Input}\SetKwInOut{Output}{Output}
\SetAlgoLined
\DontPrintSemicolon
\Input{The server $s\in\mathcal{S}$ where an arrival occurs at time $t$.}
\Fn{\FPAS{$s$}}{
	\For{$j\in\mathcal{J}$ with $s\in \mathcal{S}_j$}{
			Remove $s$ from the max heap $\mathcal{H}_j(t)$\;
			$\upsilon_j(t)\gets$ the root node of the updated $\mathcal{H}_j(t)$\;
	}
}
\caption{Updating the indication vector upon an arrival.}\label{algo:PAS:remove}

\end{algorithm}
\IncMargin{-0.5em}

For $j\in\mathcal{J}$, let
$
\mathcal{N}^{\{0\}}_{j} = \left\{ \bm{n}\in\mathcal{B} \left|\forall s\in\mathcal{S}_{j},n_{s}\in\mathcal{U}_{k(s)}\right.\right\}
$
and
$
\mathcal{N}^{\{0,1\}}_{j} = \left\{ \bm{n} \in \mathcal{B} \left|\exists s\in\mathcal{S}_{j},n_{s}\in\mathcal{C}_{k(s)}\right.\right\}$.
Rigorously, the action variables for PAS are given by \vspace{-0.2cm}
\begin{equation}\label{eqn:PAS}
a_{j,s}^{\text{PAS}}(\bm{n})=\left\{
\begin{cases}
  1,&\text{ if $\bm{n}\in\mathcal{N}_{j}^{\{0,1\}}$ and }\\
&\text{$s\in\arg\max_{s\in\mathcal{S}_{j}:n_{s}\in\mathcal{C}_{k(s)}}\Bigl[\frac{\mu_{k(s)}}{\varepsilon_{k(s)}-\varepsilon_{k(s)}^0} \Bigr]$},\\
0,&\text{ otherwise}. \vspace{-0.2cm}
\end{cases}\right.
\end{equation}
If $\arg\max[\cdot]$ returns a set with more
than one element, ties can be broken arbitrarily.  
We set, without loss of generality,
policy $\text{PAS}$ to always select the smallest $s$ among the set
of value(s) returned by this $\arg\max[\cdot]$.

For clarity, we provide an example of implementing PAS.
Maintain a \emph{indication vector} $\bm{\upsilon}(t)\in\mathcal{S}^{|\mathcal{J}|}$ at time $t$, where the $j$th element $\upsilon_j(t)$ represents the server used to accommodate a newly-arrive job of type $j$ if it arrives at time $t$. Note that, $a^{\text{PAS}}_{j,s}(\cdot)$ can be determined by $\bm{\upsilon}(t)$ by setting $a^{\text{PAS}}_{j,s}(\cdot) =1$ for $s=\upsilon_j(t)$ and $0$ for others.
For each job type $j\in\mathcal{J}$, maintain a max heap $\mathcal{H}_j(t)$ of servers $s\in\mathcal{S}_j$ with respect to the effective energy efficiency $\mu_{k(s)}/(\varepsilon_{k(s)}-\varepsilon_{k(s)}^0)$.
If a server $s$ transitions from $B_{k(s)}-1$ to $B_{k(s)}$ or from $B_{k(s)}$ to $B_{k(s)}-1$, we trigger a potential update of the indication vector and the max heaps for all types, as described in Algorithms \ref{algo:PAS:remove} and \ref{algo:PAS:add}, respectively. For both algorithms, the worst-case computational complexity is $O(\sum_{j\in\mathcal{J}}\log |\mathcal{S}_j|)$ and the space complexity is $O(\sum_{j\in\mathcal{J}}|\mathcal{S}_j|)$,
representing the storage space for the vector $\bm{\upsilon}(t)$ and heaps $\mathcal{H}_j(t)$ ($j\in\mathcal{J}$).

Note that the
$\text{PAS}$ policy does not require $\lambda_{j}$ to be known nor, indeed, do we assume specific distributions for the inter-arrival/inter-departure times. In other words,
PAS is widely applicable and scalable to a server farm with heterogeneous servers and jobs.

\DecMargin{-0.5em}
\begin{algorithm}[t]
\linespread{0.6}\selectfont
\SetKwFunction{FPAS}{UpdatingUponDeparture}
\SetKwProg{Fn}{Function}{:}{\KwRet}
\SetKwInOut{Input}{Input}\SetKwInOut{Output}{Output}
\SetAlgoLined
\DontPrintSemicolon

\Input{The server $s\in\mathcal{S}$ where a departure occurs at time $t$.}
\Fn{\FPAS{$s$}}{
	\For{$j\in\mathcal{J}$ with $s\in \mathcal{S}_j$}{
			Add $s$ in the max heap $\mathcal{H}_j(t)$\;
			$\upsilon_j(t)\gets$ the root node of the updated $\mathcal{H}_j(t)$\;
	}
}
\caption{Updating the indication vector upon a departure.}\label{algo:PAS:add}

\end{algorithm}
\DecMargin{-0.5em}

\vspace{-0.35cm}
\section{Deviation Analysis}\label{sec:deviation}
\vspace{-0.2cm}

Following similar ideas
in~\cite{weber1990index,fu2016asymptotic}, in this section, we obtain an
upper bound for the performance deviation between $\text{PAS}$ and
the optimal solution in asymptotic regime, under certain condition.  
We refer
to this upper bound as the \emph{deviation bound}.  The deviation
bound diminishes exponentially in the size of the system
leading, in particular,
to asymptotic optimality of $\text{PAS}$.

Following the idea in~\cite{whittle1988restless},  
servers should be prioritized according to their potential profits, quantized 
and obtained by relaxing
the constraint of the optimization problem.
This is referred to as the
\emph{Whittle relaxation} technique.  Our problem,
\eqref{eqn:objective}-\eqref{eqn:org:constraint_2}, is treated similarly.

This technique produces a highly intuitive heuristic scheduling policy, which coincides with PAS under certain conditions. 
We prove this equivalence in Section~\ref{subsec:relaxation}.
Based on this equivalence, in Section~\ref{subsec:performance}, we prove that the deviation bound of PAS's performance is diminishing rapidly (in fact, exponentially) as the problem size increases.

\vspace{-0.4cm}
\subsection{Whittle Relaxation}\label{subsec:relaxation}
\vspace{-0.2cm}

From~\cite[Theorem 1]{rosberg2014}, if
$\Gamma^{\phi}(\bm{f}^{\mu})<+\infty$ and
$\Gamma^{\phi}(\bm{f}^{\varepsilon})<+\infty$ for all $\phi\in\Phi$,
and we define,  \vspace{-0.1cm}
\begin{equation}\label{eqn:e_star}
e^{*} = \max\limits_{\phi\in\Phi}\left\{\Gamma^{\phi}(\bm{f}^{\mu})/\Gamma^{\phi}(\bm{f}^{\varepsilon})\right\},\vspace{-0.1cm}
\end{equation}
then a policy optimizes the problem described by equations (\ref{eqn:objective})-\eqref{eqn:org:constraint_2} if and only if it optimizes  
\begin{equation}
\label{eqn:prob_org}
\max\limits_{\phi}\left\{\Gamma^{\phi}(\bm{f}^{r})\right\},\ \text{s. t. \eqref{eqn:org:constraint_1} and \eqref{eqn:org:constraint_2},} \vspace{-0.3cm}
\end{equation}
where the vector of reward rate functions $\bm{f}^{r}\in\prod_{k\in\mathcal{K}}\mathbb{R}^{\mathcal{B}_k}$ is defined by
\begin{equation}\label{eqn:e_star:rate_func}
f^{r}_{k}(n_{k})=f^{\mu}_{k}(n_{k})-e^{*}f^{\varepsilon}_{k}(n_{k}), 
\end{equation}
for $n_k\in\mathcal{B}_{k}$, $k\in\mathcal{K}$.
Let 
\begin{equation*}
\alpha^{\phi}_{j}=\lim\limits_{t\rightarrow +\infty}\mathbb{E}[a^{\phi}_j(\bm{N}^{\phi}(t))]=\lim\limits_{t\rightarrow +\infty}\mathbb{E}\left[\Theta\left(\overline{a}^{\phi}_{j}(\bm{N}^{\phi}(t))\right)\right]
\end{equation*}
and 
\begin{equation*}
\overline{\alpha}^{\phi}_{j} = \lim\limits_{t\rightarrow+\infty}\mathbb{E}\left[\overline{a}^{\phi}_{j}(\bm{N}^{\phi}(t))\right].
\end{equation*}
The Whittle relaxation technique
involves  randomization of  the action variables
$a^{\phi}_{j,s}(\cdot)$ and $\overline{a}^{\phi}_{j}(\cdot)$.  
This relaxation of 
(\ref{eqn:objective})-\eqref{eqn:org:constraint_2}  produces the following problem: \vspace{-0.2cm}
\begin{equation}
\max\limits_{\phi} \Gamma^{\phi}(\bm{f}^{r}) \label{eqn:e_star:objective}\vspace{-0.3cm}
\end{equation}
subject to \vspace{-0.1cm}
\begin{equation}
\sum\limits_{s\in\mathcal{S}_{j}}\lim\limits_{t\rightarrow +\infty}\mathbb{E}\left[a^{\phi}_{j,s}\left(\bm{N}^{\phi}(t)\right)\right] + \alpha^{\phi}_{j}= 1,~\forall j\in\mathcal{J},\label{eqn:relax:constraint_1}
\end{equation}
\begin{equation}
\overline{\alpha}^{\phi}_{j}+ \sum\limits_{s\in\mathcal{S}_{j}}\lim\limits_{t\rightarrow +\infty}\mathbb{E}\left[I\left(B_{k(s)}-N^{\phi}_{s}(t)\right)\right] \leq 1 ,~\forall j\in\mathcal{J}.\label{eqn:relax:constraint_2}
\end{equation}
The relaxed problem no longer captures the  server farm problem
realistically, 
but is  useful for theoretical analysis.
Define
\begin{equation*}
A_{j}
=\sum_{s\in\mathcal{S}_{j}}\sum_{n\in\mathcal{C}_{k(s)}}\pi_{s}^{*}(n),~j\in\mathcal{J},
\end{equation*}
where $\pi^{*}_{s}(n)$ is the steady-state
probability of state $n$ for server $s$ under  policy $\phi$
satisfying $a^{\phi}_{j',s'}(\bm{n}')=1$ for all $\bm{n}'\in\mathcal{B}$,
$n'_{s'}\in\mathcal{C}_{k(s')}$, $j'\in\mathcal{J}$ and
$s'\in\mathcal{S}_{j'}$.  It is clear that, for $j\in\mathcal{J}$,
\begin{equation*}
A_{j} \leq \sum\limits_{s\in\mathcal{S}_{j}}\lim\limits_{t\rightarrow +\infty}\mathbb{E}\left[I\left(B_{k(s)}-N^{\phi}_{s}(t)\right)\right].
\end{equation*}
In this context, Equation~\eqref{eqn:relax:constraint_2} can   be
relaxed further to
\begin{equation}\label{eqn:relax:constraint_3}
\Theta(\overline{\alpha}_{j}^{\phi}) \leq \Theta(1-A_{j}) ,\ \forall j\in\mathcal{J}.
\end{equation}

To complete the analysis, we introduce, for the relaxed problem defined by \eqref{eqn:e_star:objective},\eqref{eqn:relax:constraint_1} and \eqref{eqn:relax:constraint_3},  a redundant constraint:
\begin{equation}\label{eqn:relax:constraint_4}
\lim\limits_{t\rightarrow\infty}\mathbb{E}\bigl[a^{\phi}_{j,s}(\bm{N}^{\phi}(t))|\ N^{\phi}_{s}=B_{k(s)}\bigr]=0,
\left\{
\begin{array}{l}
\forall j\in\mathcal{J},\\ 
\forall s\in\mathcal{S}_{j},
\end{array}\right.
\end{equation}
which forces the action varibles for the uncontrollable states to be zero.
Constraints~\eqref{eqn:relax:constraint_1},
\eqref{eqn:relax:constraint_3} and \eqref{eqn:relax:constraint_4} can be
combined with the objective function by introducing Lagrangian
multipliers
$\bm{\nu}$,
$\bm{\gamma}$ and
$\bm{\eta}$
 corresponding to \eqref{eqn:relax:constraint_1}, \eqref{eqn:relax:constraint_3} and \eqref{eqn:relax:constraint_4}, respectively.
Let $\tilde{\Phi}$ represent the set of all stationary policies.
Define
\begin{enumerate}
\item $\pi_{s}^{\phi}(n)$, $s\in\mathcal{S}$, $n\in\mathcal{B}_{k(s)}$, $\phi\in\tilde{\Phi}$, as the steady state probability of server $s$ in state $n$ under policy $\phi$;
\item row vector $\bm{\pi}^{\phi}_{s}=(\pi^{\phi}_{s}(n):\ n\in\mathcal{B}_{k(s)})$, $s\in\mathcal{S}$, $\phi\in\tilde{\Phi}$;
\item column vector $\bm{f}^{r}_{s}=(f^{r}_{s}(n):\ n\in\mathcal{B}_{k(s)})$, $s\in\mathcal{S}$;
\item column vector $\bm{\alpha}^{\phi}_{j,s}=(\alpha^{\phi}_{j,s}(n):\ n\in\mathcal{B}_{k(s)})$ where 
\begin{equation*}
\alpha^{\phi}_{j,s}(n) = \lim\limits_{t\rightarrow +\infty}\mathbb{E}\left[a^{\phi}_{j,s}(\bm{N}^{\phi}(t))|\ N^{\phi}_{s}(t)=n\right],
\end{equation*}
$j\in\mathcal{J}$, $s\in\mathcal{S}_{j}$, $\phi\in\tilde{\Phi}$;
\item column vector $\bm{e}^{n}$, $n\in\mathbb{N}^{+}$, of size $n$ with all zero elements except the $n$th element set to be one.
\end{enumerate}
The Lagrange problem with respect to the primal problem defined by \eqref{eqn:e_star:objective}, \eqref{eqn:relax:constraint_1}, \eqref{eqn:relax:constraint_3} and \eqref{eqn:relax:constraint_4} is then 
\begin{multline}\label{eqn:dual}
\Lambda(\bm{\nu},\bm{\gamma},\bm{\eta})=\\
\max\limits_{\phi}
\sum\limits_{s\in\mathcal{S}}\bm{\pi}_{s}^{\phi}\Bigl(\bm{f}^{r}_{s} - \sum\limits_{j\in\mathcal{J}_{s}}\nu_j\bm{\alpha}^{\phi}_{j,s}  -\sum\limits_{j\in\mathcal{J}_{s}}\eta_{j,s}\alpha^{\phi}_{j,s}(B_{k(s)})\bm{e}^{B_{k(s)}}\Bigr)\\
-\sum\limits_{j\in\mathcal{J}}(\nu_{j}\alpha^{\phi}_{j}+\gamma_{j}\Theta(\overline{\alpha}^{\phi}_{j}))+\sum\limits_{j\in\mathcal{J}}\left(\nu_{j}+\gamma_{j}\Theta(1-A_{j})\right)
\end{multline}
where  
$\mathcal{J}_{s}=\{j\in\mathcal{J}|\ s\in\mathcal{S}_{j}\}$, $s\in\mathcal{S}$.

As in~\cite{whittle1988restless},   given $\bm{\nu}$, $\bm{\gamma}$
and $\bm{\eta}$, the maximization problem at the right hand
side of \eqref{eqn:dual} achieves the same maximum as a sum of the maximum values of $|\mathcal{S}|+J$ independent sub-problems:
for $s\in\mathcal{S}$, \vspace{-0.2cm}
\begin{equation}\label{eqn:sub_problem:1}
\max\limits_{\phi\in\Phi_{s}}\bm{\pi}_{s}^{\phi}\bigl( \bm{f}^{r}_{s} - \sum\limits_{j\in\mathcal{J}_{s}}\nu_j\bm{\alpha}^{\phi}_{j,s}
- \sum\limits_{j\in\mathcal{J}_{s}}\eta_{j,s}\alpha^{\phi}_{j,s}(B_{k(s)})\bm{e}^{B_{k(s)}}\bigr)
\vspace{-0.2cm}
\end{equation}
where $\Phi_{s}$ represents the set of stationary policies $\phi$ determined by action variables $\alpha^{\phi}_{j,s}(n)\in[0,1]$, $n\in\mathcal{B}_{k(s)}$, $j\in\mathcal{J}$; and 
for $j\in\mathcal{J}$, \vspace{-0.2cm}
\begin{equation}\label{eqn:sub_problem:2}
\max\limits_{\phi\in\overline{\Phi}_{j}}-\nu_{j}\alpha^{\phi}_{j}-\gamma_{j}\Theta(\overline{\alpha}^{\phi}_{j}), \vspace{-0.2cm}
\end{equation}
where $\overline{\Phi}_{j}$ is the set of stationary policies $\phi$ determined by action variables $\overline{\alpha}^{\phi}_{j}$ and $\alpha^{\phi}_{j}$.
Remarkably, the dimension of the state space for each of these independent sub-problems is $1$.

\vspace{-0.2cm}
\begin{condition}\label{cond:heavy_traffic}
For all $j\in\mathcal{J}$, $A_{j}\leq 1$.\vspace{-0.2cm}
\end{condition}
We refer to Condition~\ref{cond:heavy_traffic} as the \emph{heavy
  traffic condition}: the blocking probabilities for jobs are \emph{almost}
positive all the time.\vspace{-0.2cm}
\begin{proposition}\label{prop:opt_equal}
When the job sizes are exponentially distributed, 
if either $J=1$ or Condition~\ref{cond:heavy_traffic} holds true, then
there exists a $\nu\in \mathbb{R}$ and 
a policy $\phi^{*}\in\Phi$ that maximizes the relaxed problem defined by  \eqref{eqn:e_star:objective}, \eqref{eqn:relax:constraint_1}, \eqref{eqn:relax:constraint_3} and \eqref{eqn:relax:constraint_4}, satisfying, for $j\in\mathcal{J}$, \vspace{-0.2cm}
\begin{enumerate}
\item if $s\in\mathcal{S}_{j}$, $n\in\mathcal{C}_{k(s)}$, \vspace{-0.3cm}
\begin{equation}\label{eqn:prop:opt_equal:1}
\alpha^{\phi^{*}}_{j,s}(n)=\left\{\begin{cases}
1, & \text{if } \nu < 1-e^{*}\frac{\varepsilon_{k(s)}-\varepsilon_{k(s)}^{0}}{\mu_{k(s)}}\,\\
1\text{ or }0, & \text{if }\nu = 1-e^{*}\frac{\varepsilon_{k(s)}-\varepsilon_{k(s)}^{0}}{\mu_{k(s)}},\\
0, & \text{if } \nu > 1-e^{*}\frac{\varepsilon_{k(s)}-\varepsilon_{k(s)}^{0}}{\mu_{k(s)}};
\end{cases}\right.
\end{equation}
\item and 
\begin{equation}\label{eqn:prop:opt_equal:2}
\alpha^{\phi^{*}}_{j}=1-\sum\limits_{s\in\mathcal{S}_{j}}\sum\limits_{n\in\mathcal{C}_{k(s)}}\pi^{\phi^{*}}_{s}(n)\alpha^{\phi^{*}}_{j,s}(n).
\end{equation}
\end{enumerate}
\end{proposition}
The proof of Proposition~\ref{prop:opt_equal} is given in Appendix~\ref{app:prop:opt_equal}.

Note that although we can calculate the optimal solution of the relaxed problem, referred to as policy $\phi^*$ in Proposition~\ref{prop:opt_equal}, such $\phi^*$ is not applicable to the original problem that we are really interested in.
Nevertheless, $\phi^*$ does offer interesting intuitions that help us construct a scalable, near-optimal heuristic policy applicable to the original problem. 
In the following, we explain and construct this heuristic policy and  prove its equivalence to PAS.

For $j\in\mathcal{J}$, $s\in\mathcal{S}_{j}$,
$n\in\mathcal{C}_{k(s)}$, 
let\vspace{-0.3cm}
\begin{equation}\label{eqn:index}
\nu^{*}_{j,k(s)}(n)= 1-e^{*}\frac{\varepsilon_{k(s)}-\varepsilon_{k(s)}^{0}}{\mu_{k(s)}}.
\end{equation}

The optimal policy $\phi^*$ described in \eqref{eqn:prop:opt_equal:1}
implies the priorities of different servers: for a given $\nu$, if job-server pair $(j,s)$ has $\alpha^{\phi^*}_{j,s}(n)=1$, $n\in\mathcal{C}_{k(s)}$, then any other pairs $(j,s')$ with $\nu^{*}_{j,k(s')}(n')>\nu^{*}_{j,k(s)}(n)$ must have $\alpha^{\phi^*}_{j,s'}(n')=1$, $n'\in\mathcal{C}_{k(s')}$.
The value of $\nu^{*}_{j,k(s)}(\cdot)$ can be interpreted as the server's potential profits  (or subsidy following the idea in \cite{whittle1988restless}) gained by choosing this server $s$ to serve a job of type $j$ if this server is not fully occupied.

A similar property was defined in \cite{whittle1988restless} for a Restless Multi-Armed Bandit Problem (RMABP) and referred to as \emph{Whittle indexability}.
When job sizes are exponentially distributed, $J=1$ and neglecting \eqref{eqn:org:constraint_2}, our problem reduces to a 
RMABP, and Whittle indices are given by $\nu^{*}_{j,k(s)}(\cdot)$.  
There may be no simple closed form for the  Whittle indices in  the general case with general job-size distributions.  
Relevant work about RMABP and Whittle indices have been mentioned in Section~\ref{sec:rWork}.

In the general case, if we always assign incoming jobs of type $j$ to servers with the highest $\nu^{*}_{j,k(s)}(\cdot)$, for the highest potential profits, among those in the available set $\mathcal{S}_j$ and with vacancies in their buffers,
then the resulting policy coincides with PAS described in Section~\ref{sec:index}. 
PAS is applicable to the original problem,
defined by
\eqref{eqn:objective}-\eqref{eqn:org:constraint_2}. 
We discuss PAS in Section~\ref{subsec:performance} by comparing it to policy $\phi^*$ described in Proposition~\ref{prop:opt_equal}.
If $\phi^*$ is optimal for the relaxed problem, then the performance of $\phi^*$ is an upper bound of that of the original problem. If PAS's performance again coincides with $\phi^*$, then PAS is optimal for the original problem.

\vspace{-0.3cm}
\subsection{Convergence in Performance}\label{subsec:performance}
\subsubsection{Stochastic Processes with Smooth Trajectories}\label{subsubsec:asym_opt}

Let states
$n\in \mathcal{C}_k$ for all $k\in\mathcal{K}$ be
ordered according to descending values $\nu^{*}_{j,k}(n)$, where uncontrollable
states $n\in\mathcal{U}_k$, for all $k\in\mathcal{K}$, follow the controllable
states in the ordering, with $\alpha^{\phi}_{j,s}(n)=0$ for
$n\in\mathcal{U}_k$, $s\in\mathcal{R}_k$, $k\in\mathcal{K}_j$, $j\in\mathcal{J}$.
Then we place the state $n\in \mathcal{B}_{0}$ of zero-reward
servers, also a controllable state, after all other
controllable states but preceding the uncontrollable states.  
To indicate the order of states, the position of a state in the ordering $i=1,2,\ldots,I$, where
$I=\sum_{k\in\mathcal{K}\cup\{0\}}|\mathcal{B}_{k}|$,
is regarded as  its label.  
Define
$\tilde{\mathcal{B}}
\coloneqq\{1,2,\ldots, I\}$.
Let $n_i$ represent the server state labeled by $i$ (i.e., the $i$th state), and $k_i$ represent the only server group with $n_i\in\mathcal{B}_{k_i}$.

Since each $i\in\tilde{\mathcal{B}}$ is associated with a
   server group $k$ and a state in $\mathcal{B}_k$, servers are thus
 distinguishable only through their current state $i\in\tilde{\mathcal{B}}$.

Let
$\bm{Z}^{\phi}(t)=(Z^{\phi}_{i}(t):\ i\in\tilde{\mathcal{B}})$, 
and $\mathcal{Z}$ represent the space of all probability vectors of
length $I$. 
The random variable $Z_{i}^{\phi}(t)$ represents the proportion of
servers in state $i\in\tilde{\mathcal{B}}$ at time $t$ under policy
$\phi\in\Phi$, $t\geq 0$: that is,
\begin{equation*}
Z_{i}^{\phi}(t) \coloneqq \frac{1}{|\mathcal{S}|+R_{0}}
\left|\left\{s\in\mathcal{S}\ \right.\left|\ N_{s}^{\phi}(t) = i\right\}\right|.
\end{equation*}
Define a mapping $m$ by $m(\bm{N}^{\phi}(t)) = \bm{Z}^{\phi}(t)$. Recall that our server farm is assumed to be empty at $t=0$, and, correspondingly, define $\bm{z}^0\coloneqq\bm{Z}^{\phi}(0)=m(\bm{0})$.
On the arrival or departure of jobs at time $t$, $\bm{Z}^{\phi}(t)$
transitions to $\bm{Z}^{\phi}(t)+ \bm{e}_{i,i'}$, where
$\bm{e}_{i,i'}$ is a vector of which the $i$th element is
$+1/(|\mathcal{S}|+R_{0})$, the $i'$th element is
$-1/(|\mathcal{S}|+R_{0})$ and otherwise is zero,
$i,i'\in\tilde{\mathcal{B}}$.  Servers in server group $k$ only
appear in state $n_i\in\mathcal{B}_{k}$; that is, the transition from
$\bm{Z}^{\phi}(t)$ to $\bm{Z}^{\phi}(t)+ \bm{e}_{i,i'}$,
$n_i\in \mathcal{B}_{k}$, $n_{i'}\in\mathcal{B}_{k'}$,
$k,k'\in\mathcal{K}\cup\{0\}$, $k\neq k'$ never occurs.

Let $R_{k} = R_{k}^{0}h$
and
$\lambda_{j}=\lambda_{j}^{0}h$, 
where $h=1,2,\ldots$ is called the \emph{scaling parameter}. Correspondingly, let
$S_0 \coloneqq(|\mathcal{S}|+R_{0})/h$, then $S_0\in\mathbb{N}^{+}$.
Since servers in the same state $i\in\tilde{\mathcal{B}}$ are
indistinguishable, we define the probability of selecting/activating a server
in state $i$ for an arriving job of type $j$, i.e., the probability of
setting $\alpha^{\phi}_{j,s}(i)=1$, as
$u^{\phi,h}_{j, i}(\bm{z})$, when policy $\phi$ is used,  and $\bm{Z}^{\phi}(t)=\bm{z}$.    Clearly,
$u^{\phi,h}_{j,i}(\bm{z})=0$ and
$\bm{z}\in\mathcal{Z}$, if $i$ represents an uncontrollable state.

We obtain, for $i\in\tilde{\mathcal{B}}$,
$j\in\mathcal{J}$, $h\in\mathbb{N}^{+}$, $\bm{z}\in\mathcal{Z}$ with
$z_{i}>0$,
\begin{equation*}\label{eqn:probability_active}
u^{\text{PAS},h}_{j,i}(\bm{z}) = 
\min\biggl\{1,\frac{1}{z_{i}}\max\Bigl\{0,\frac{1}{hS_0}-\sum\limits_{i'<{i},k_{i'}\in\mathcal{K}_{j}}z_{i'}\Bigr\}\biggr\}.
\end{equation*}

We define without loss of generality the transition caused by an
arrival event in state $i$ as the transition from
state $i$ to state $i+1$, if $n_i,n_{i+1}\in\mathcal{C}_{k}$,
$k\in\mathcal{K}$.  
Such a transition from $i$ to $i+1$ is caused by
an  arrival of a job of a specified type.

Following the ideas of~\cite{weber1990index,fu2016asymptotic,fu2018restless}, we then obtain a corollary of~\cite[Proposition 5]{fu2018restless} as follows. 
\begin{corollary}\label{coro:equilibrium_vector}
When the job sizes are exponentially distributed, 
for any $\delta>0$,  there exists a $\bm{z}^{\text{PAS}}\in\mathcal{Z}$ such that
\begin{equation}\label{eqn:equilibrium_vector}
\lim\limits_{h\rightarrow +\infty}\lim\limits_{t\rightarrow +\infty} \frac{1}{t}\int_{0}^{t}\mathbb{P}\Bigl\{\bigl\lVert\bm{Z}^{\text{PAS},h}(u)-\bm{z}^{\text{PAS}}\bigr\rVert>\delta\Bigr\}du = 0,
\end{equation}
with given $\bm{Z}^{\text{PAS},h}(0)=\bm{z}^0$.
\end{corollary}

Equation~\eqref{eqn:equilibrium_vector} indicates that stochastic process $\bm{Z}^{\text{PAS},h}(t)$ will go into a close neighborhood of point $\bm{z}^{\text{PAS}}$ as the scaling parameter $h$ tends to infinity, where the process transition rates of leaving and entering each of its states must be equivalent. Also, since priorities of states are driven by $u^{\text{PAS},h}_{j,i}(\cdot)$,  if we start from an empty server farm, the trajectory of $Z^{\text{PAS},h}_{i}(t)$, $t\geq 0$, is independent from the trajectories of $Z^{\text{PAS},h}_{i'}(t)$, $t\geq 0$, for states $i' > i$ (that is, states with lower priorities). We can then calculate the value of $\bm{z}^{\text{PAS}}$ from the first element to the last, which coincides with the calculating procedure of $\lim_{h\rightarrow +\infty}\lim_{t\rightarrow +\infty}\mathbb{E}[\bm{Z}^{\phi^*,h}(t)]$ under $\phi^*$ with the state priorities driven by action variables described in \eqref{eqn:prop:opt_equal:1} and \eqref{eqn:prop:opt_equal:2}.

Given the reward rate functions for all states $i\in\tilde{\mathcal{B}}$, the long-run average reward $\Gamma^{\phi}(\bm{f}^r)$ (the objective function of our problem) of policy $\phi$ is linear in the expected value $\lim_{t\rightarrow +\infty} \mathbb{E}\bm{Z}^{\phi,h}(t)$.
In other words, when  $\bm{Z}^{\text{PAS},h}(t)$ approaches $\bm{z}^{\text{PAS}}$ in the asymptotic regime,
$\Gamma^{\text{PAS}}(\bm{f}^{r})$ also approaches  $\Gamma^{\phi^*}(\bm{f}^{r})$ for any given $e^*\in\mathbb{R}$.
As mentioned at the end of Section~\ref{subsec:relaxation}, 
if $\phi^*$ is  also  optimal for the relaxed problem, then PAS is asymptotically optimal in the original problem, because the maximized energy efficiency of the relaxed problem is always an upper bound of that of the original one.

In consequence, if $J=1$ or Condition~\ref{cond:heavy_traffic} is
satisfied, Proposition~\ref{prop:opt_equal} and
Corollary~\ref{coro:equilibrium_vector} yield the asymptotic
optimality of $\text{PAS}$ as $h\rightarrow +\infty$ in terms of
energy efficiency, when  job sizes are exponentially distributed.

\subsubsection{Bounded Performance Deviation}\label{subsubsec:deviation}

According to our result on  asymptotic optimality stated in
Section~\ref{subsubsec:asym_opt}, the performance deviation between
$\text{PAS}$ and the optimal solution in the asymptotic regime is directly related to the
supremum of Euclidean distances between
$\lim_{t\rightarrow +\infty}\mathbb{E}\bm{Z}^{\text{PAS},h}(t)$ and
$\bm{z}^{\text{PAS}}$.

\begin{proposition}\label{prop:sup_equilibrium}
When the job sizes are exponentially distributed, 
for any $\delta>0$, there exist $\bm{z}^{\text{PAS}}\in\mathcal{Z}$, $s>0$ and $H>0$, such that,
for any $h > H$, \vspace{-0.2cm}
\begin{equation}\label{eqn:sup_equilibrium_3}
\lim\limits_{t\rightarrow +\infty} \frac{1}{t}\int_{0}^{t}\mathbb{P}\Bigl\{\bigl\lVert\bm{Z}^{\text{PAS},h}(u)-\bm{z}^{\text{PAS}}\bigr\rVert>\delta\Bigr\}du \leq e^{-sh}, \vspace{-0.2cm}
\end{equation}
where $\bm{Z}^{\text{PAS},h}(0)=\bm{z}^0$.
\end{proposition}
The proof of Proposition~\ref{prop:sup_equilibrium} is given in Appendix~\ref{app:prop:sup_equilibrium}.
That is, the deviation bound of PAS diminishes exponentially in the scaling parameter $h$.

As mentioned in Section~\ref{sec:introduction}, since the asymptotic regime can never be achieved in the real world, Proposition~\ref{prop:sup_equilibrium} sharpens the asymptotic optimality result: PAS approaches asymptotic optimality very quickly (exponentially) as the problem size increases.
Simulation results will be provided in Section~\ref{sec:numerical}.

Proposition~\ref{prop:sup_equilibrium} applies in systems with much more general power functions, 
extending asymptotic optimality of PAS to more general cases. 
For instance, by \cite[Proposition 1]{fu2018restless} and Proposition~\ref{prop:sup_equilibrium}, the PAS family is also asymptotically optimal when the service and energy consumption rates of each server is linearly increasing in the number of jobs there, although the linearity is not appropriate in modeling power consumption in Cloud environments. As mentioned in Section~\ref{sec:rWork}, the problem studied 
here is already sufficiently complex to prevent existing methods from being applied directly.
More general power functions will presumably and significantly complicate the system model and notational definitions and is outside the scope of this paper.

\begin{figure}[t]
\centering
\subfigure[]{\includegraphics[width=0.43\linewidth]{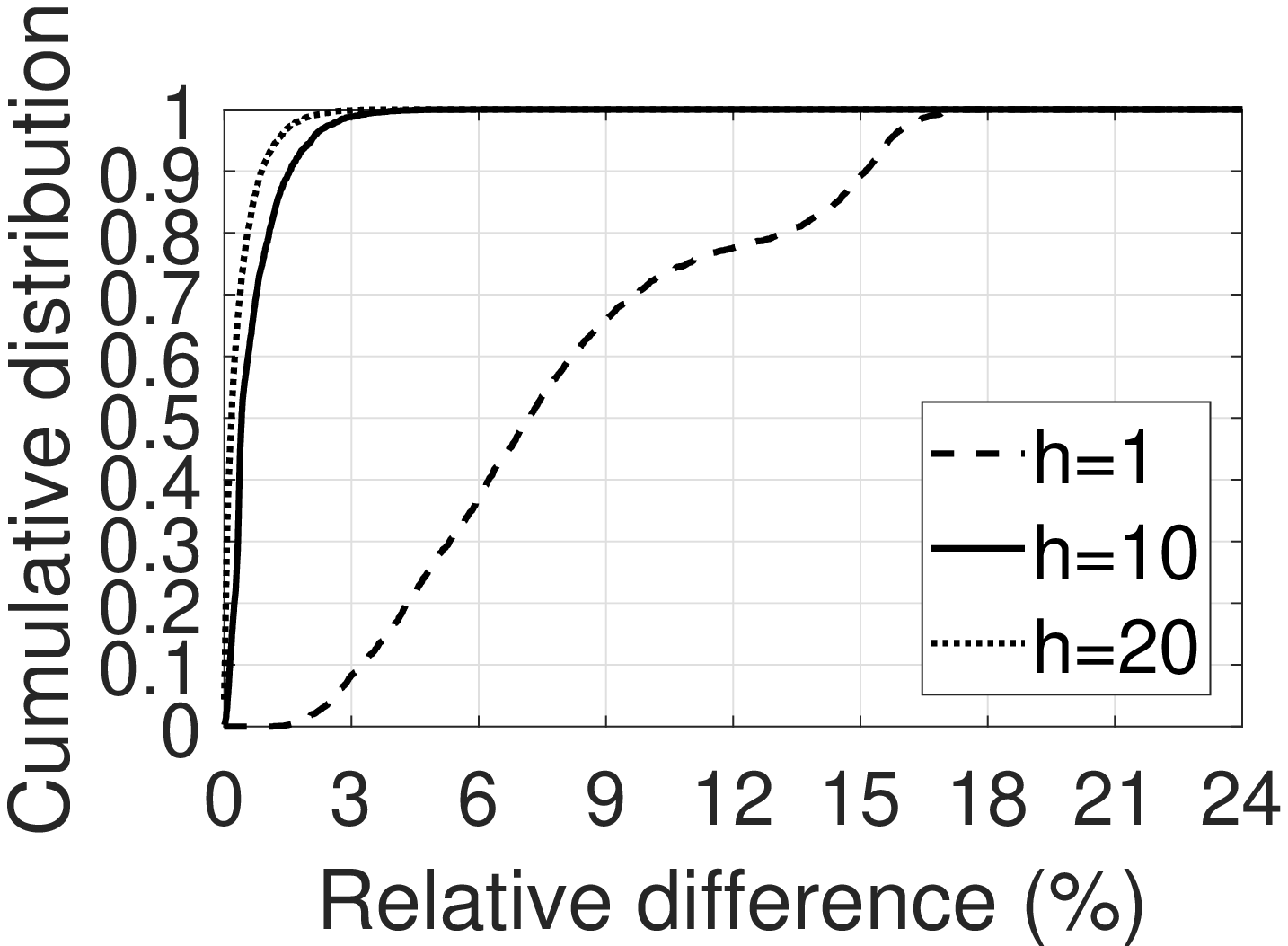}\label{fig:fig1a}}
\subfigure[]{\includegraphics[width=0.43\linewidth]{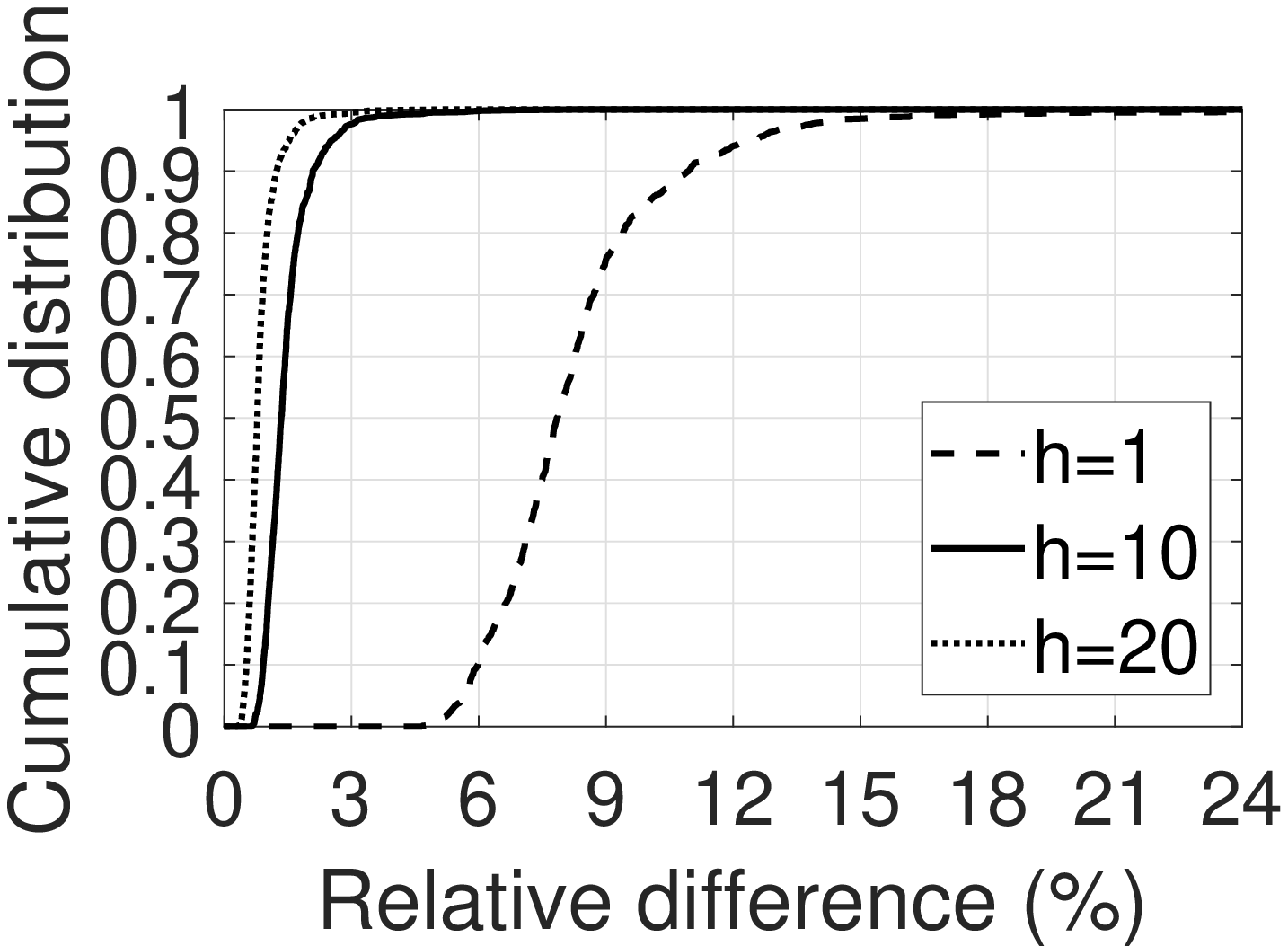}\label{fig:fig1b}}
\caption{Cumulative distribution of the normalized performance deviation of the PAS policy: (a) single job type; (b) multiple job types.}\label{fig:fig1}
\end{figure}

\vspace{-0.3cm}
\section{Numerical Results}\label{sec:numerical}

Here, we numerically demonstrate the effectiveness of PAS by comparing it with the optimal energy efficiency, as a benchmark, in different scenarios with randomly generated server farm systems. We recall that, as mentioned in Sections~\ref{sec:introduction} and \ref{sec:rWork}, the complexity of the server farm model prevents applicability of existing scheduling policies from being direct.

 In our simulation results, the  $95\%$
confidence intervals based on the Student $t$-distribution are
maintained within $\pm 3\%$ of the observed mean.  
In Sections~\ref{sec:deviation_simulation} and \ref{subsec:google}, we set job-sizes to be
exponentially distributed, and in
Section~\ref{subsec:sensitivity}, more realistic job-size
distributions are discussed.

\vspace{-0.3cm}
\subsection{Deviation Studies}\label{sec:deviation_simulation}
We are interested in the performance deviation of the PAS policy; that
is, how large the system should be to guarantee a performance deviation with reasonable bounds. For the sake of simplicity, let
OPT represent an optimal solution for the relaxed problem defined by
\eqref{eqn:e_star:objective}, \eqref{eqn:relax:constraint_1},
\eqref{eqn:relax:constraint_3} and \eqref{eqn:relax:constraint_4} in the asymptotic regime, we
define the \emph{normalized performance deviation} of a policy
$\phi\in\Phi$, to be
\begin{equation*}
\Bigl(\mathcal{L}^{\rm OPT}/\mathcal{E}^{\rm OPT}-\mathcal{L}^{\phi}/\mathcal{E}^{\phi}\Bigr)/\Bigl(\mathcal{L}^{\rm OPT}/\mathcal{E}^{\rm OPT}\Bigr).
\end{equation*}
Note that energy efficiency under OPT is an upper bound for that
under an optimal solution of the original problem defined by
\eqref{eqn:objective}--\eqref{eqn:org:constraint_2} in the asymptotic regime.
Because of the
extremely high computational complexity of the original problem, we
use OPT as a benchmark in our numerical experiments.

\subsubsection{Stochastically Identical Jobs}\label{subsec:one_job}
We start with the simple case of only one job type ($J=1$).  Consider
a server farm with five server groups ($K=5$), each of which has
$R_k^0=1$, $k\in\mathcal{K}$, servers when the scaling parameter
$h=1$. Let the buffer sizes $B_k$  of all servers equal 2,
for all $k\in\mathcal{K}$.  In Figure~\ref{fig:fig1a}, we depict
the cumulative distribution of the normalized performance deviation of
PAS in terms of energy efficiency, with randomly generated service
rates, energy consumption rates and sets of available servers as
follows.
\begin{itemize}
\item Service rates $\mu_k$, $k\in\mathcal{K}$, are randomly uniformly
  generated in the range $[1,10]$;
\item Energy efficiencies of servers in the first group are normalized
  to be 1, i.e., $\mu_1/\varepsilon_1=1$; those in successive groups
  are obtained by randomly uniformly generating the  ratio of server energy efficiencies for successive
  groups, i.e., $(\mu_k/\varepsilon_k)/(\mu_{k-1}/\varepsilon_{k-1})$,
  $k=2,3,\ldots,K$, from $[0.5,1]$ iteratively;
\item According to the service rates and energy efficiencies of servers
  for different groups, we obtain the busy power consumption for all
  servers, and set the idle power of servers in group $k$,
  $k\in\mathcal{K}$, to be $\varepsilon_k\cdot(0.1+0.1\cdot k)$;
\item All servers in the server farm are available for incoming jobs
  $\mathcal{K}_1=\mathcal{K}$; and 
\item The average arrival rate $\lambda_1$ is set to $\rho\cdot \sum_{k=1}^{K}\mu_{k}$ with a given \emph{normalized offered traffic} $\rho$.
\end{itemize}

In Figure~\ref{fig:fig1a}, the value of the normalized performance deviation of the
PAS policy is decreasing in $h$, $h=1,10,20$, and, for all our simulations, is within $3\%$ (of the  energy efficiency under OPT) when $h=20$; that is,   
twenty
servers in each server group.  In other words, in this experiment,
the PAS policy is already close to OPT when the
server farm is relatively small.  These results
are consistent with the deviation upper bound described by
\eqref{eqn:sup_equilibrium_3}.  
That is, PAS is demonstrated to be near-optimal since the scaling parameter is relatively small, 
and, in line with our theoretical results \eqref{eqn:sup_equilibrium_3}, is likely to be near optimal for any larger $h$.

\begin{figure*}[t]
\centering
\begin{minipage}[]{0.24\textwidth}
\includegraphics[width=\linewidth]{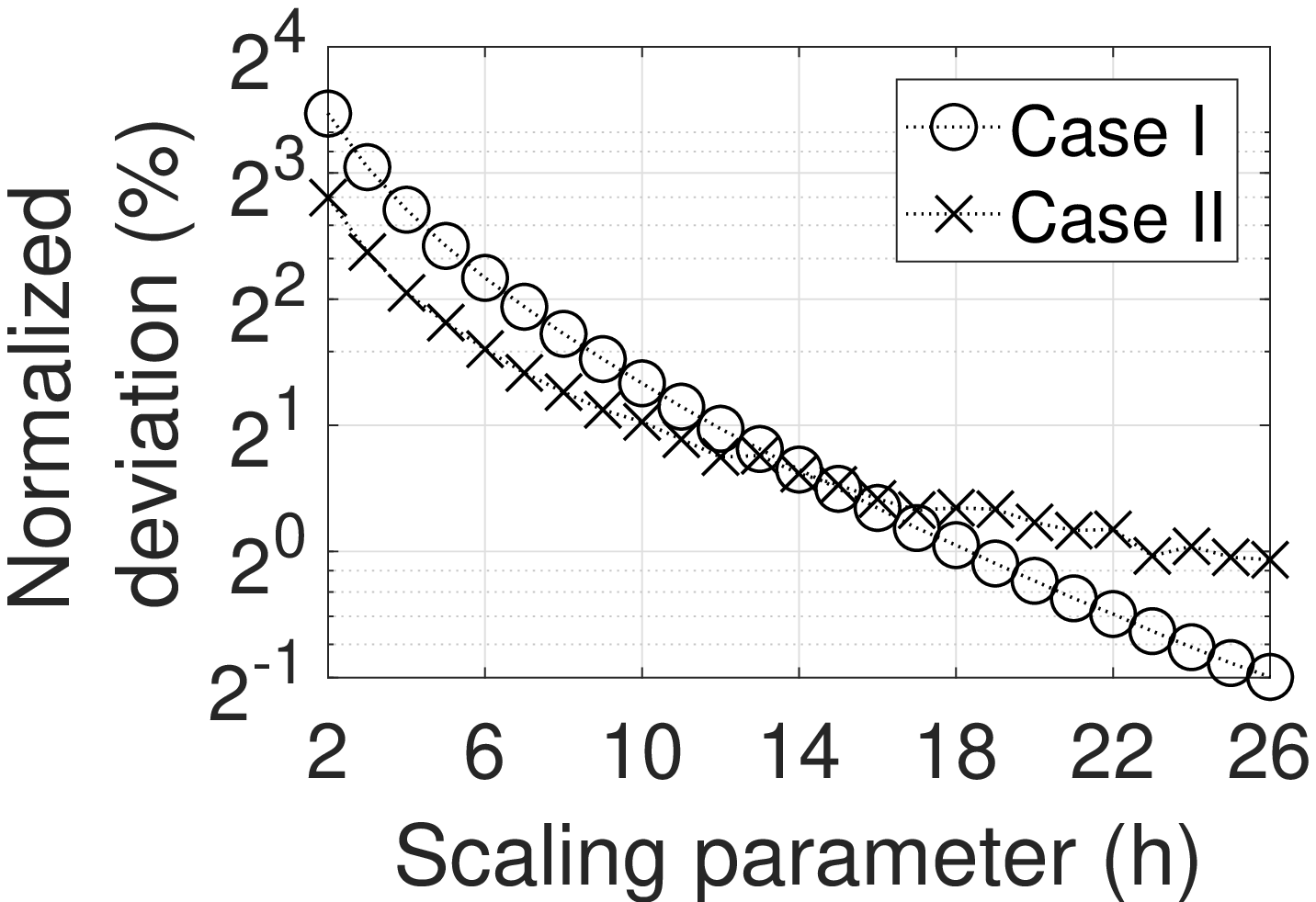}\caption{Performance of the PAS policy against the scaling parameter. \label{fig:fig2}}
\end{minipage}
\begin{minipage}[]{0.24\textwidth}
\includegraphics[width=\linewidth]{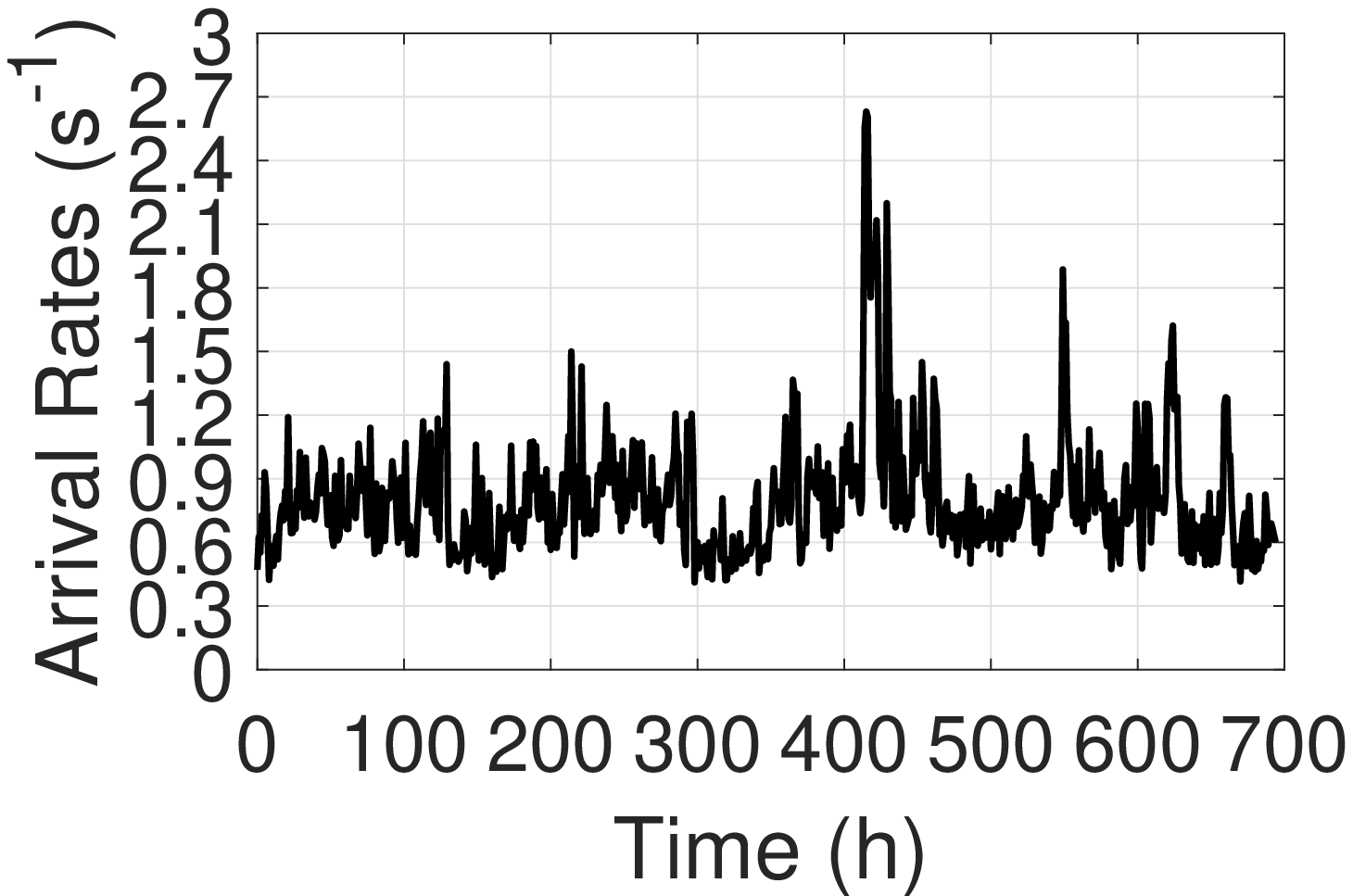}
\caption{Arrival rates of Google trace-logs.\label{fig:google-arrival}}
\end{minipage}
\begin{minipage}[]{0.24\textwidth}
\includegraphics[width=\linewidth]{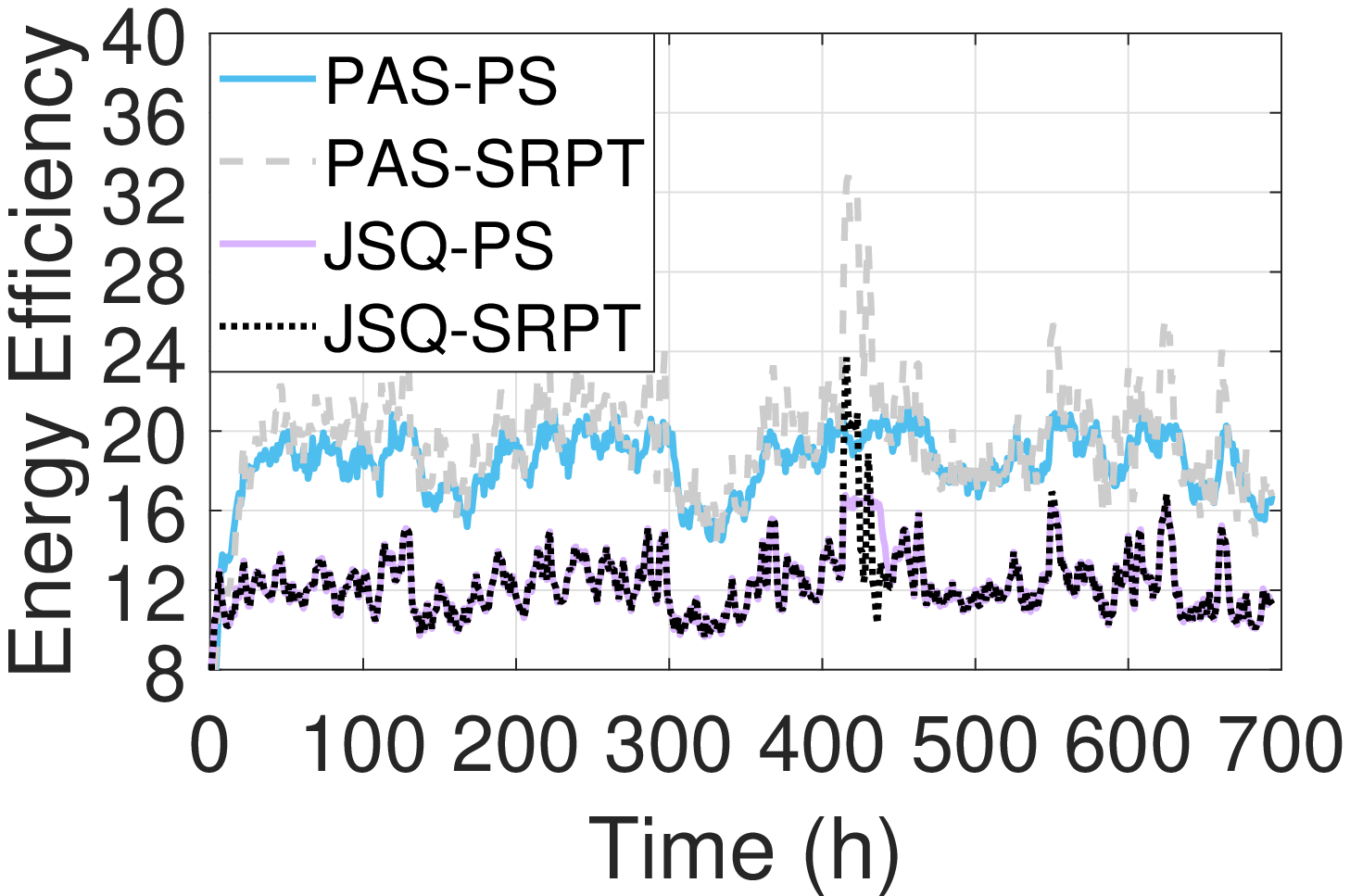}
\caption{Energy efficiency of PAS and JSQ.\label{fig:google-TE}}
\end{minipage}
\begin{minipage}[]{0.24\textwidth}
\includegraphics[width=\linewidth]{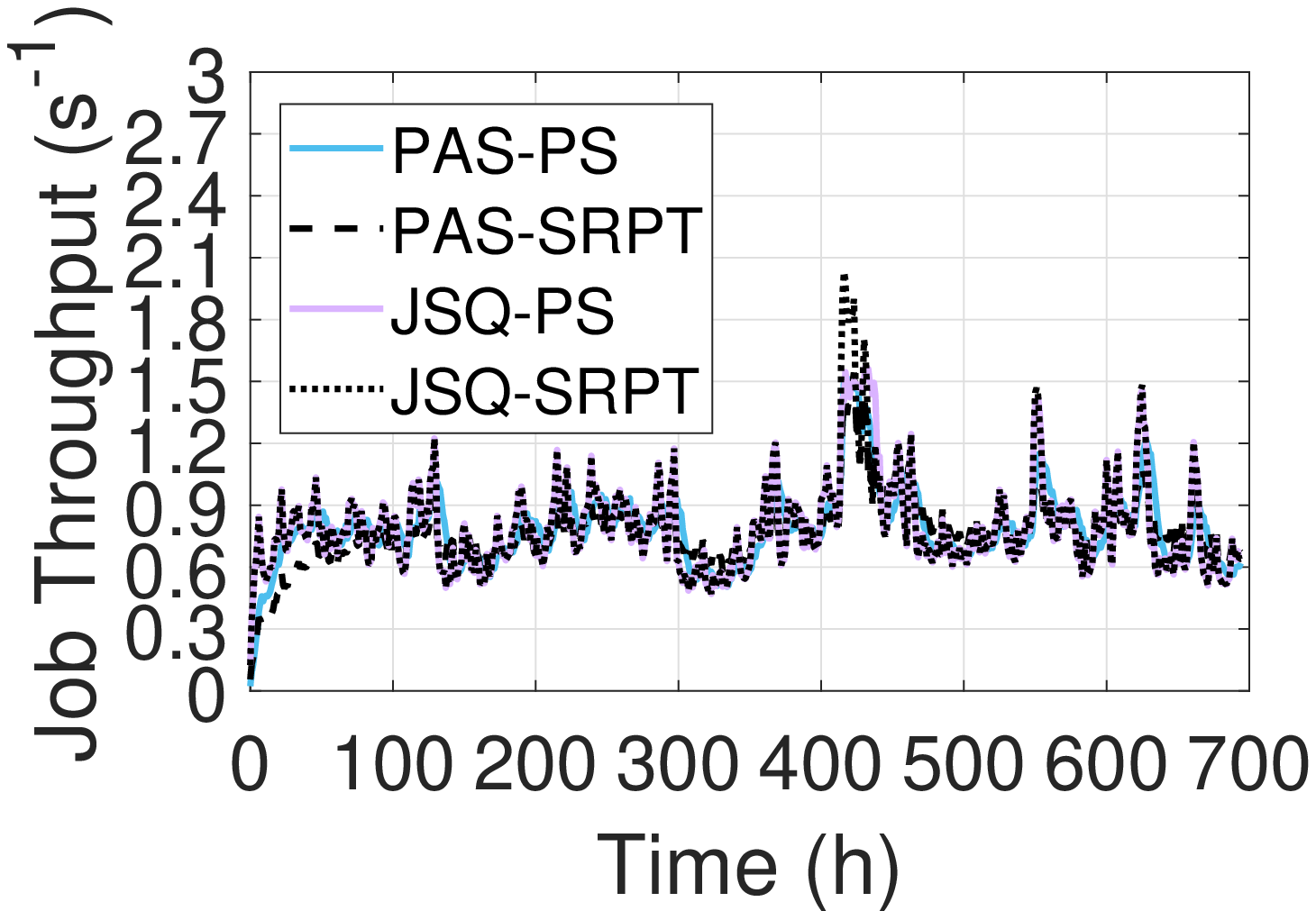}
\caption{Job throughput of PAS and JSQ.\label{fig:google-throughput}}
\end{minipage}
\vspace{-0.5cm}
\end{figure*}

\subsubsection{Multiple Job Types}\label{subsec:multi_job}
For the general case of multiple job types, we define servers with the
same settings as those for Figure~\ref{fig:fig1a} except here we set
$B_k=1$ for $k\in\mathcal{K}$.  We plot the cumulative distribution of
normalized performance deviation of the PAS policy in
Figure~\ref{fig:fig1b}, where three different job types have been
considered ($J=3$).  The parameters for a job of type 
$j\in\mathcal{J}$, are generated by:
\begin{itemize}
\item we firstly  generate a  random number $m_j$ of server groups for
  available servers, following a uniform distribution within $\{1,2,\ldots,K\}$;
\item then we randomly pick  $m_j$ server groups from the total $K$
  ones as the server groups for available servers, and generate the
  set $\mathcal{K}_j$ of these server groups;
\item set the average arrival rate of jobs of type $j$ to be the
  product of $\rho$ and the sum of service rates of all its available
  servers, where $\rho$ is given.
\end{itemize}
We compare PAS with OPT in Figure~\ref{fig:fig1b} where the heavy
traffic condition (Condition~\ref{cond:heavy_traffic}) is satisfied,
so that an optimal solution for the relaxed problem (OPT) exists in
the form of \eqref{eqn:prop:opt_equal:1} and
\eqref{eqn:prop:opt_equal:2}.  Note that this heavy traffic condition
is not necessary for the $J=1$ case discussed in
Section~\ref{subsec:one_job}.  For the general case with $J >1$, if the heavy traffic condition is not valid,
OPT does not necessarily exist in the form of
\eqref{eqn:prop:opt_equal:1} and \eqref{eqn:prop:opt_equal:2} and it
remains unclear how to calculate such an OPT within a reasonable time.

In Figure~\ref{fig:fig1b}, the normalized performance deviation of
PAS maintains a trend similar to that in Figure~\ref{fig:fig1a},
decreasing quickly with increasing $h$, $h=1,10,20$.  
The normalized performance deviation of PAS is no more than $3\%$ in
almost all experiments for Figure~\ref{fig:fig1b} when $h=20$. This is
consistent with our argument regarding  Figure~\ref{fig:fig1a} that PAS is
close to OPT even for a relatively small system
and so for such a system with any larger scaling parameter $h$.

We pick up two specific runs of simulations for
Figures~\ref{fig:fig1a} and \ref{fig:fig1b} as examples for both cases,
and, in Figure~\ref{fig:fig2}, demonstrate 
the normalized performance deviation of PAS against the scaling
parameter $h$ of the server farm for two cases: \emph{Case I} and \emph{II}  stand for systems with stochastically identical jobs and multiple job types, respectively. 
The
detailed parameter values, which are generated randomly, 
for Figure~\ref{fig:fig2} are provided in Appendix~\ref{app:fig2}.

In Figure~\ref{fig:fig2}, the normalized deviation of PAS is seen to
approach 0 as $h$ increases, being greater than
 $1\%$ for  $h\geq26$, for both cases. 
Figure~\ref{fig:fig2} has a plot of the normalized performance deviation of PAS
against the scaling parameter $h$, with the $y$-axis in  log
scale.  The curve for Case I appears almost linear in
$h$, and that for Case II  convex in $h$, with an almost linear
tail.  These results are consistent with the exponentially
decreasing upper bound of PAS performance deviation described by
\eqref{eqn:sup_equilibrium_3}.  The straight curve for Case I and the
straight tail for Case II suggest that the upper bound shown in
\eqref{eqn:sup_equilibrium_3} is likely to be  tight.

All the demonstrated simulations with randomly generated parameters have shown convergence between PAS and OPT in energy efficiency since the scaling parameter $h$ is relatively small, implying the near-optimality of PAS for any larger $h$. PAS is thus appropriate for server farms with realistic scales; that is,  large but not necessarily in the asymptotic regime.

\vspace{-0.3cm}
\subsection{Case Studies}\label{subsec:google}
\vspace{-0.2cm}
We now consider the performance of PAS with respect to Google cluster traces of job arrivals in 2011 \cite{clusterdata:Wilkes2011, clusterdata:Reiss2011}. The cluster consists of 12.5 thousand machines with arriving jobs classified into four groups. The job arrival rates, estimated as the number of arrived jobs per second, averaged in each hour are plotted in Figure~\ref{fig:google-arrival}.

In Figures~\ref{fig:google-TE} and \ref{fig:google-throughput}, we demonstrate the effectiveness of the PAS policy by comparing it with a baseline policy, Join-the-Shortest-Queue (JSQ).  JSQ is a load balancing policy that is proved to maximize the number of processed jobs within a given time period \cite{winston1977optimality}.

Unlike in the simulations presented in Section~\ref{sec:deviation_simulation}, in this subsection, we do not assume Poisson arrival process, so that
the different service disciplines potentially lead to different steady state distributions or the long-run average performance of the system. 
Consider two classical service disciplines for the simulation results in this subsection and Section~\ref{subsec:sensitivity}: the PS and
the Shortest-Remaining-Processing-Time (SRPT)  disciplines.
The PS discipline of processing jobs is appropriate for web server farms to avoid unfair processing delays between jobs, especially when their job sizes are highly varied \cite{gupta2007analysis, altman2011load}.
The SRPT is a well-known  discipline that minimizes the mean response time \cite{bansal2001analysis}.

There are ten server groups each of which contains 1.25 thousand servers with randomly generated service and power consumption rates and availability for serving jobs of different types.
The detailed parameter values for simulations in this subsection are provided in Appendix~\ref{app:fig_google}.

Observing Figures~\ref{fig:google-TE} and \ref{fig:google-throughput}, for either service discipline, PAS achieves clearly higher energy efficiency while maintaining comparable job throughput with those of JSQ. 
For the PAS policy, 
the energy efficiency and job throughput curves for SRPT are slightly higher than those for PS in terms of both energy efficiency and job throughput. 
This is because SRPT is a discipline aiming at load balancing while PS is designed for guaranteeing fairness and robustness.

Moreover, for the simulations in Figures~\ref{fig:google-TE} and \ref{fig:google-throughput}, although the traffic intensity during the peak hours is higher than one (heavy traffic condition is satisfied), the simulated number of blocked jobs is zero for both PAS and JSQ. Because the scale of the entire server farm is sufficiently large, with $12.5$ thousand servers, and the peak hours with heavy traffic are relatively few as shown in Figure~\ref{fig:google-arrival},  there are sufficiently many buffer slots in the server farm to digest the heavy traffic during peak hours and thus
the number of blocked jobs is negligible in the presented simulations.

\begin{figure}[t]
\centering
\begin{minipage}[]{0.225\textwidth}
\includegraphics[width=\linewidth]{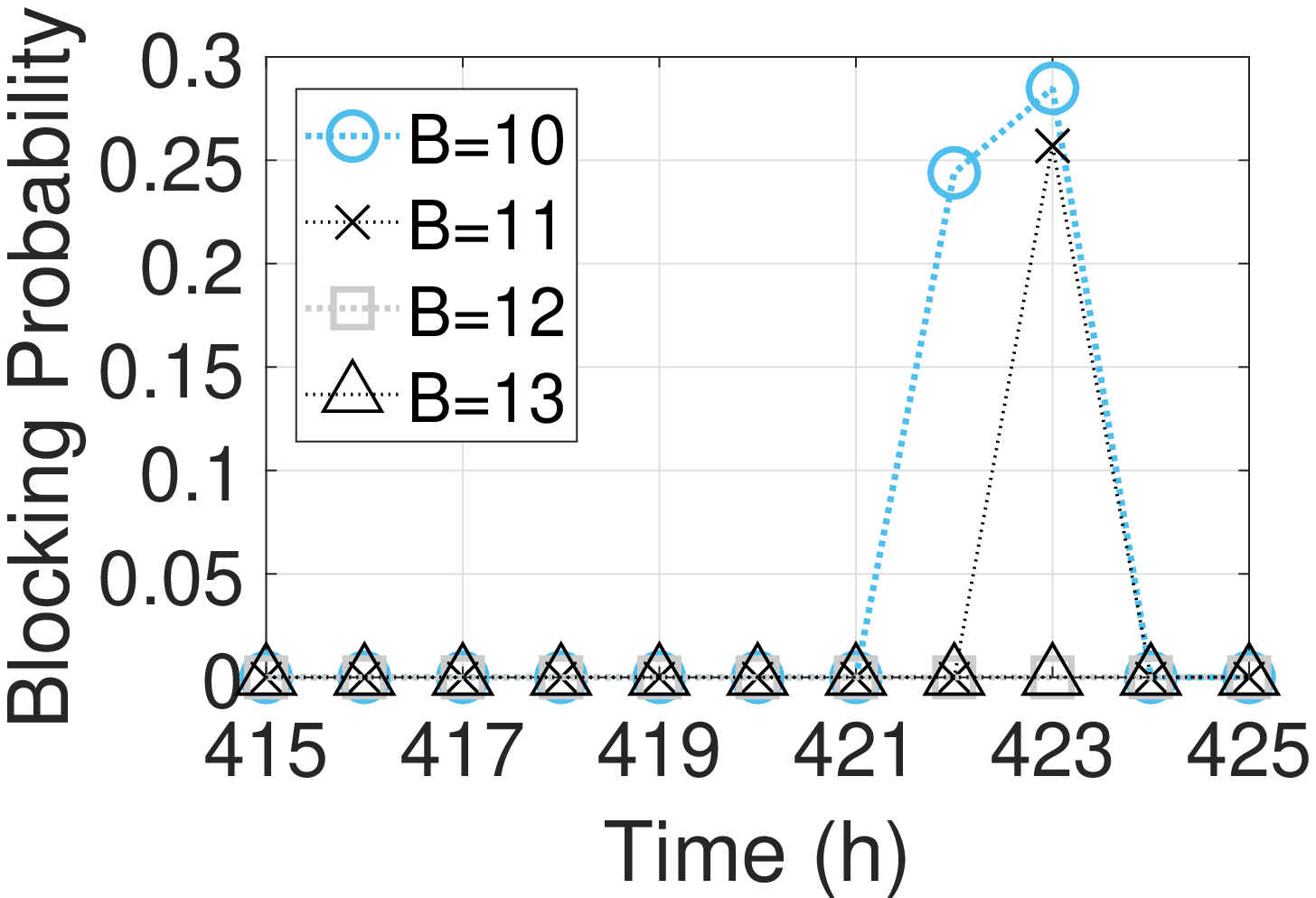}
\caption{Blocking probability of PAS under PS during the peak hours.\label{fig:google-blocking}}
\end{minipage}
\begin{minipage}[]{0.225\textwidth}
\includegraphics[width=\linewidth]{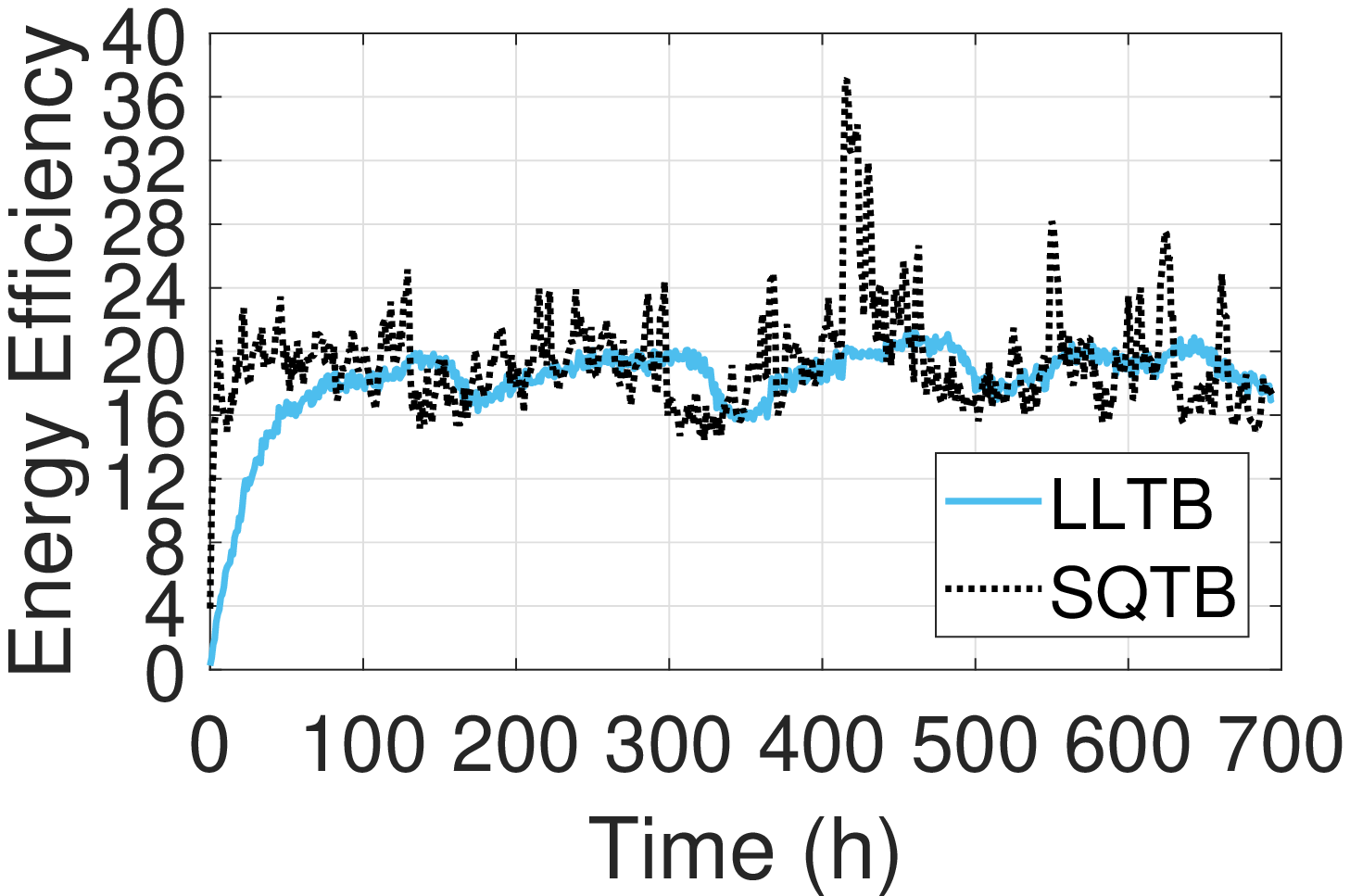}
\caption{Energy efficiency of PAS with different tie-breaking rules.\label{fig:google-tie}}
\end{minipage}
\end{figure}

Also, we tested the blocking probability of the Google trace-logs used in Figure~\ref{fig:google-throughput}. The system parameters are the same as those for the simulations presented in Figure~\ref{fig:google-throughput}, except that the server buffer sizes are set to $B$ and the $B$ takes different values: $10,11,12,13$. For all the tested buffer sizes, PAS under SRPT and JSQ under both disciplines incur zero blocked jobs based on our simulations; while, as demonstrated in Figure~\ref{fig:google-blocking}, PAS under PS incurs non-negligible blocking probabilities  during the peak hours for $B=10,11$ and this reduces to zero for $B\geq 12$. 
These results strengthen our earlier argument:  when the total buffer size of the entire server farm is sufficiently large, the number of blocked jobs becomes negligible even if the heavy traffic condition is satisfied. For a large server farm with 12.5 thousand servers, such as the Google cluster mentioned above, the total buffer size of the entire server farm is already large with relative small $B$.

Moreover,  to complete the discussion, in Figure~\ref{fig:google-tie}, we plot the energy efficiency of PAS under PS with a different tie-breaking rule, but the same settings as the simulations presented in Figure~\ref{fig:google-throughput}.
Recall that all our theoretical results apply to any tie-breaking rule, and, as described in Section~\ref{sec:index}, we have just chosen the simplest: when there is more than one server with the highest effective energy efficiency, assign jobs to the server with lowest label. We refer to it as \emph{Lowest Label Tie Breaking (LLTB)}.
Alternatively, for multiple servers with the same effective energy efficiency, we could choose the one with the shortest queue; this is referred to as \emph{Shortest Queue Tie Breaking (SQTB)}.

We can see from Figure~\ref{fig:google-tie} that LLTB achieves  flatter performance, while SQTB is more variable. But in reality there appears to be no general advantage of one over the other. 
In particular, SQTB outperforms LLTB by around $0.5\%$ with respect to the total energy efficiency and both cases incur no blocked jobs.
We then argue that PAS under PS is not very sensitive to different tie-breaking rules. 
The robustness of PAS to service disciplines demands further exploration involving more comprehensive case studies with real-world trace-logs, but that is outside the scope of this paper. 

\vspace{-0.2cm}
\subsection{Robustness Studies}\label{subsec:sensitivity}
\vspace{-0.2cm}

In practice, job duration times for many online applications have
been studied and known to be 
characterized by heavy tailed distributions~\cite{harchol2013performance}, which is at odds with our exponential assumption. Hence, 
it is important to understand the sensitivity of the PAS policy to different
job-size distributions.  In this context, we consider two heavy-tailed
distributions: 
Pareto with shape parameter $2.001$ (Pareto-F for short) and
Pareto with shape parameter $1.98$ (Pareto-INF for
short); these are  set to have unit mean.
Note that Pareto-F and Pareto-INF are Pareto distributions with finite and infinite variance, respectively.

With the same settings
as for Section~\ref{subsec:multi_job}, here, we test the energy efficiency of PAS with exponentially, Pareto-F and Pareto-INF distributed job sizes.
We also consider a case where the job sizes of different types are distributed differently; 
this is referred to as the \emph{mixed} case. 

\begin{figure}[t]
\vspace{-0.3cm}
\centering
\subfigure[]{\includegraphics[width=0.44\linewidth]{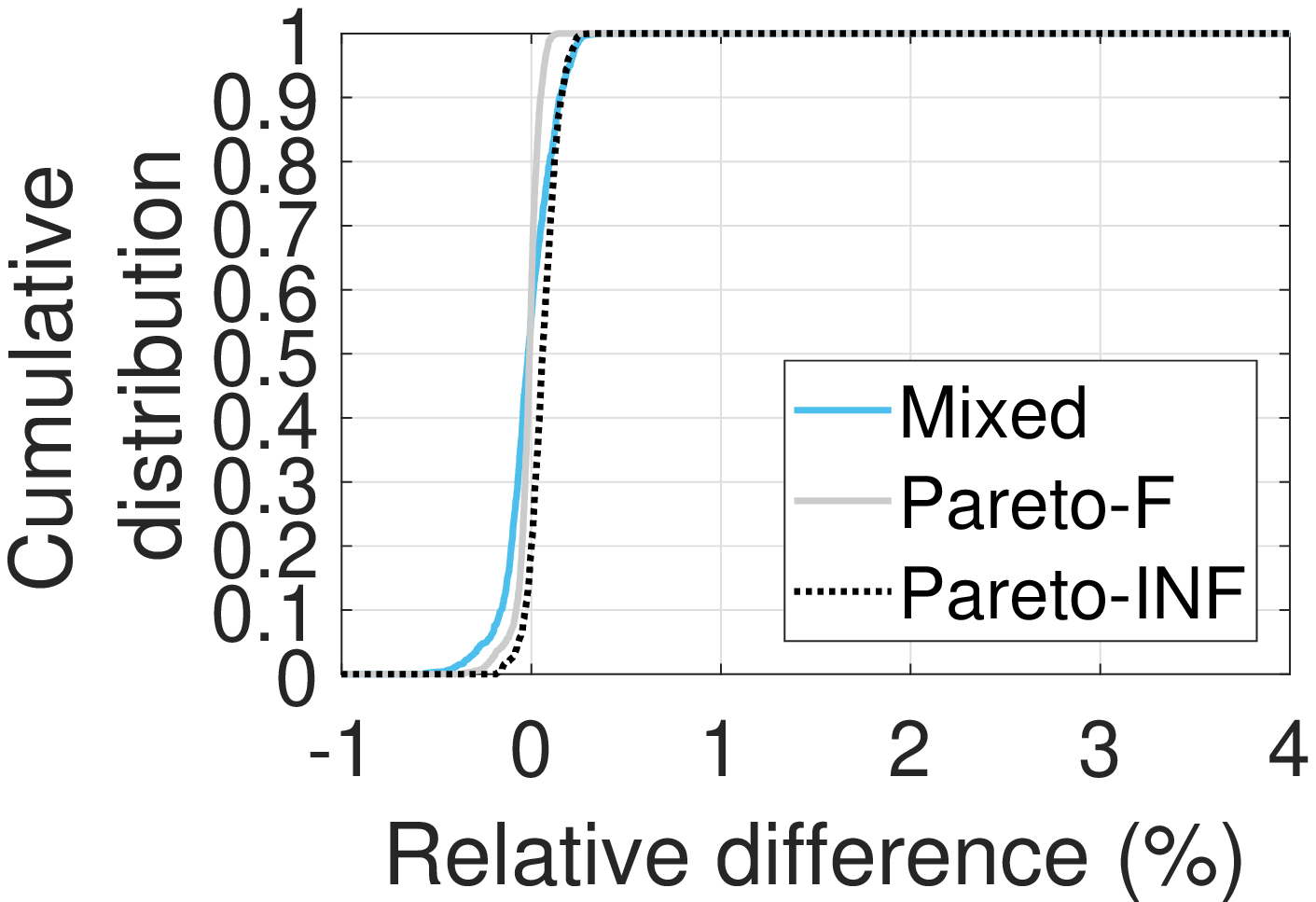}\label{fig:fig5a}}
\subfigure[]{\includegraphics[width=0.44\linewidth]{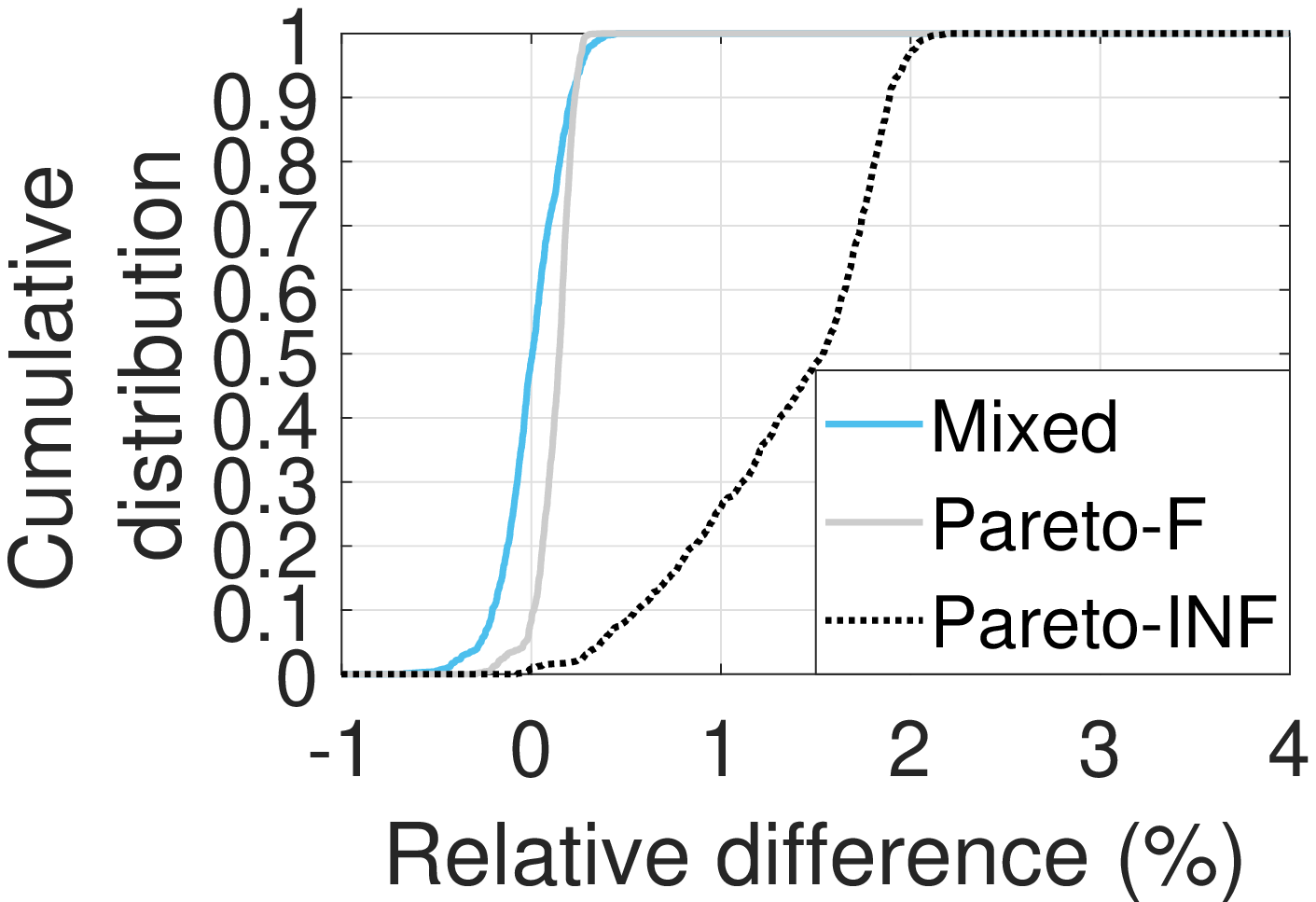}\label{fig:fig5b}}
\vspace{-0.3cm}
\caption{Cumulative distribution of the relative difference of $\Gamma^{D}$ to $\Gamma^{\rm exponential}$ with $D$ = mixed, Pareto-F or Pareto-INF under (a) the PS discipline; and (b) the SRPT discipline. }\label{fig:fig5}
\end{figure}

In Figure~\ref{fig:fig5}, we demonstrate the robustness of PAS under the PS and SRPT disciplines with respect to different job size distributions.
Let $\Gamma^{D}$ represent the energy efficiency of the server farm
under PAS with job-size distribution $D$, where $D=$ exponential, mixed, Pareto-F or Pareto-INF.

In Figure~\ref{fig:fig5}, we show the cumulative distribution of
$(\Gamma^{D}-\Gamma^{\rm exponential})/\Gamma^{\rm exponential}$; that is, the
 relative difference of energy efficiency with  job size distribution $D$  from  the one with exponentially distributed job sizes.
In Figure~\ref{fig:fig5a}, this relative difference is within $\pm 1\%$ in all our experiments with randomly generated parameters; while, varies between $-1\%$ and $3\%$ in Figure~\ref{fig:fig5b}. 
PAS is resilient to these tested job-sizes distributions under PS and SRPT, although SRPT incurs slightly higher variance than PS.

\vspace{-0.3cm}
\section{Conclusions}\label{sec:conclusion}

We have studied the job-assignment problem in a server farm
model consisting of a large number of abstracted servers 
that are possibly
diverse in 
service rates, energy consumption rates, buffer sizes (service capacities) and
the ability to serve different jobs.  
Also, as described in Section~\ref{sec:model}, in this work, the relationship between energy consumption and service rates of servers (abstracted computer components) can be arbitrary, and be  determined by the functional features and profiles of servers.
By assigning jobs to efficient
servers, we aim to balance the job throughput and the power consumption of the system; that is, we aim to maximize the energy efficiency defined as
the ratio of long-run average job departure rate to the long-run
average energy consumption rate of the entire server farm system.

Following the idea of
Whittle relaxation~\cite{whittle1988restless}, we have proposed the
scalable  
policy, PAS,  that  prioritizes servers according to only
intrinsic attributes and binary state information of different
servers 
 among the available ones. 
PAS accounts for the availability of servers to service different jobs, enabling the applicability of our server farm model to
geographically separated computing systems.

To the best of our knowledge, there is no existing work that proposes scalable, infinite horizon policies for such heterogeneous server farms at realistic scales, with a rigorous analysis of performance deviation in terms of energy efficiency.

We have proved that, when job sizes are
exponentially distributed, if the blocking probabilities of jobs are
always positive or $J=1$, there exists a deviation bound for PAS which
is  exponentially decreasing  in the number of servers in server groups
and the average arrival rates of jobs proportionately.  This deviation
bound indicates the asymptotic optimality of PAS, and,
more importantly, significantly improves the asymptotic optimality results:
PAS approaches asymptotic optimality very quickly (exponentially) as the server farm size increases.
Numerical results  illustrate that PAS is already close
to OPT for  only  100 servers,
 consistent with our  deviation bound. We infer that PAS is
nearly optimal for even a relatively small system and any larger one.  The
robustness of PAS to three different job-size distributions has
been tested numerically, with resulting values of energy
efficiency in our simulations  close to those with exponentially
distributed job sizes.

\appendices
\vspace{-0.3cm}
\section{Proof of Proposition~\ref{prop:opt_equal}}\label{app:prop:opt_equal}

Fix
a server,  $s\in\mathcal{S}$, starting in state
$n_s\in\mathcal{B}_{k(s)}$, a policy $\phi\in
  \Phi_s$, and a reward rate $f_{k(s)}\in\mathbb{R}^{\mathcal{B}_{k(s)}}$.  Write $V^{\phi}_{s}(n_{s},f_{k(s)})$,
 for the expected value of the cumulative reward for server $s$ that
 ends when it first enters an absorbing state $n^{0}_{s}\in\mathcal{B}_{k(s)}$.
In particular,
$V^{\phi}_{s}(n^{0}_{s},f_{k(s)}) = 0$ for any $\phi\in\Phi_s$.  We
assume, without loss of generality, that $n^{0}_{s}=0$ for all
$s\in\mathcal{S}$.  Note that such a $V^{\phi}_{s}(n_s,f_{k(s)})$ is
dependent on the values of $\bm{\nu}$ and $\bm{\gamma}$, though we
refrain from including them  as superscript/subscript to simplify
notation. 
Let
$V_{s}(n_{s},f_{k(s)}) =
\max_{\phi\in\Phi_{s}}V^{\phi}_{s}(n_{s},f_{k(s)})$,
$f_{k(s)}\in\mathcal{R}_{k(s)}$, $s\in\mathcal{S}$,
$n_s\in\mathcal{B}_{k(s)}$.

Now, let $\mathcal{P}^{H}_{s}$, $s\in\mathcal{S}$, represent a process for server $s$ that starts from state $0$ until it reaches state $0$ again, where $\phi\in\Phi_s$ is constrained to those policies satisfying $\sum_{j\in\mathcal{J}_{s}}\alpha_{j,s}^{\phi}(0)>0$.
From  \cite[Corollary 6.20 and Theorem 7.5]{ross1992applied}, when the job sizes are exponentially distributed, the process $\mathcal{P}^{H}_{s}$ is a renewal interval of the long-run process, so that the average reward of $\mathcal{P}^{H}_{s}$ equals the long-run average reward of the process for server $s$ under the same policy.

Define, for $s\in\mathcal{S}$, $\phi\in\Phi_s$, $n\in\mathcal{B}_{k(s)}$,
\begin{displaymath}
\tilde{f}_{k(s)}(n)=\left\{\begin{cases}
f^{r}_{k(s)}(n)-\sum\limits_{j\in\mathcal{J}_{s}}\nu_{j}\alpha^{\phi}_{j,s}(n), & \text{if } n\in\mathcal{C}_{k(s)},\\
f^{r}_{k(s)}(n)-\sum\limits_{j\in\mathcal{J}_{s}}(\nu_{j}+\eta_{j,s})\alpha^{\phi}_{j,s}(n), & \text{otherwise},
\end{cases}\right.
\end{displaymath}
where $f^{r}_{k(s)}(n)$ is as defined in \eqref{eqn:e_star:rate_func}.
Following \cite[Theorem 7.6, Theorem 7.7]{ross1992applied} and \cite[Corollary 1]{fu2016asymptotic}, 
there exists $g\in\mathbb{R}$, with
$\tilde{f}^{g}_{k(s)}(n)=\tilde{f}_{k(s)}(n)-g$,
$n\in\mathcal{B}_{k(s)}$,  such that if policy
$\phi^{*}\in\Phi_{s}^{H}$ maximizes the expected cumulative reward of
process $\mathcal{P}^{H}_{s}$ with reward rate
$\tilde{f}^{g}_{k(s)}(n)$, then $\phi^{*}$ also maximizes the long-run
average reward of server $s$ with reward rate $\tilde{f}_{k(s)}(n)$ among all policies in $\Phi_{s}^{H}$.  This value of $g$, denoted by $g^{*}_{s}$, is just  the maximized long-run average reward.

For $s\in\mathcal{S}$, $j\in\mathcal{J}_{s}$,
$n\in\mathcal{C}_{k(s)}$ again for notational simplicity,  we write
$V^{g}_{s}(n)=V_{s}(n,\tilde{f}^{g}_{k(s)})$ and
$r^{g}_s(n)=f^{r}_{k(s)}(n)-g$. Note that, in this context, $V^{g}_{s}(0)=V_{s}(0,\tilde{f}^{g}_{k(s)})=0$ since state $0$ is the absorbing state.

\begin{lemma}\label{lemma:independent_j}
When the job
  sizes are exponentially distributed, there exists an policy
  $\phi^{*}\in\Phi_{s}$, $s\in\mathcal{S}$, that maximizes the
  objective function given by \eqref{eqn:sub_problem:1}, satisfying:
  for $n\in\mathcal{C}_{k(s)}$,
\begin{equation}\label{eqn:lemma:independent_j:1}
\alpha^{\phi^{*}}_{j,s}(n)=\left\{\begin{cases}
1, & \text{if } \frac{\nu_{j}}{\lambda_{j}} < V_{s}^{g^*}(n+1) - V_{s}^{g^*}(n),\\
1\text{ or } 0, & \text{if }\frac{\nu_{j}}{\lambda_{j}} = V_{s}^{g^*}(n+1) - V_{s}^{g^*}(n),\\
0, & \text{otherwise},
\end{cases}\right.
\end{equation}
where $g^{*}\coloneqq g^{*}_{s}$ for notational simplicity.
\end{lemma}

\begin{IEEEproof}
We start from \eqref{eqn:lemma:independent_j:1}.
Let 
$\overline{\lambda}_{j,s}(n) = \sum_{j'\in\mathcal{J}_{s}:\ j\neq j',\ a^{\phi^{*}}_{j',s}(n)=1}\lambda_{j'}$.
Then, from the Bellman equation for  Markov Decision Process (MDP), we obtain \eqref{eqn:lemma:independent_j:1} for $n\in\mathcal{C}_{k(s)}\backslash\{0\}$,
which takes values independent of  $\overline{\lambda}_{j,s}(n)$ and identically for all $j\in\mathcal{J}_{s}$.

Similarly, the maximization problem based on the Bellman equation for state $n=0$ lead to, for $\overline{\lambda}_{j,s}(0)>0$,
equation~\eqref{eqn:lemma:independent_j:1} when $g$ is set to be $g^{*}$.
For $\overline{\lambda}_{j,s}(0)=0$, there exists an optimal policy $\phi^{*}\in\Phi_s$ with $a^{\phi^{*}}_{j,s}(0)=1$ if and only if 
$V^{g^{*}}_{s}(1)-(e^{*}\varepsilon_{k(s)}^{0}+\nu_{j}+g^{*})/\lambda_{j} = 0 $,
and $g^{*} \geq -e^{*}\varepsilon_{k(s)}^{0}$,
since $g^{*}$ equals the optimal average reward of the same $\mathcal{P}^{H}_{s}$ with reward rate $\tilde{f}_{k(s)}(n)$, $n\in\mathcal{B}_{k(s)}$.
These two equations lead to \eqref{eqn:lemma:independent_j:1} with $n=0$.
The lemma is proved.
\end{IEEEproof}

\begin{lemma}\label{lemma:opt_ratio}
When the job sizes are exponentially distributed,
if $\nu = \nu_{j}/\lambda_{j}$ for all $j\in\mathcal{J}$, there exists
a policy $\phi^{*}\in\Phi_{s}$ ($s\in\mathcal{S}$) maximizing the
objective function given by \eqref{eqn:sub_problem:1}, and satisfying \eqref{eqn:prop:opt_equal:1} for $s\in\mathcal{S}_{j}$ and $n\in\mathcal{C}_{k(s)}$.
\end{lemma}
\begin{IEEEproof}
From Lemma~\ref{lemma:independent_j}, for any $s\in\mathcal{S}_{j}$,
the maximization problem defined in \eqref{eqn:sub_problem:1} is equivalent to
\begin{equation*}
\max\limits_{m\in\mathcal{C}_{k(s)}}\lambda\biggl(\frac{\mu_{k(s)}-e^{*}(\varepsilon_{k(s)}-\varepsilon_{k(s)}^{0})}{\mu_{k(s)}}-\nu\biggr)\frac{\sum\limits_{n=0}^{m}\left(\frac{\lambda}{\mu_{k(s)}}\right)^{n}}{\sum\limits_{n=0}^{m+1}\left(\frac{\lambda}{\mu_{k(s)}}\right)^{n}},
\end{equation*}
where $\lambda = \sum_{j\in\mathcal{J}_{s}}\lambda_{j}$. Solving it, we obtain \eqref{eqn:prop:opt_equal:1}.
\end{IEEEproof}

We now give the proof of Proposition~\ref{prop:opt_equal}. 
\begin{IEEEproof}
The proof consists of two parts: 
\begin{enumerate}[label=$(\alpha*)$]
\item we construct a set of $\bm{\nu}$, $\bm{\gamma}$, $\varphi_j\in\overline{\Phi}_{j}$, $j\in\mathcal{J}$, and $\phi^{*}_{s}\in\Phi_{s}$, $s\in\mathcal{S}$, where $\phi^{*}_{s}$ and $\varphi_j$ maximize the objective functions defined in \eqref{eqn:sub_problem:1} and \eqref{eqn:sub_problem:2}, respectively; 
\item and we prove that a policy consisting of such $\varphi_j$, $j\in\mathcal{J}$, and $\phi^{*}_{s}$, $s\in\mathcal{S}$, maximizes the relaxed problem defined by \eqref{eqn:e_star:objective}, \eqref{eqn:relax:constraint_1}, \eqref{eqn:relax:constraint_3}-\eqref{eqn:relax:constraint_4}.
\end{enumerate}
We firstly discuss the case where Condition~\ref{cond:heavy_traffic} holds.
Let 
\begin{enumerate}
\item $\nu_{j}/\lambda_{j} = \nu$ and $\nu =\min\Bigl\{0, \min\limits_{s\in\mathcal{S}_{j}}\bigl(1-e^{*}\frac{\varepsilon_{k(s)}-\varepsilon_{k(s)}^{0}}{\mu_{k(s)}}\bigr)\Bigr\}$; 
\item $\gamma_{j} = - \lambda_{j}\nu$, $j\in\mathcal{J}$.
\end{enumerate}
From Lemma~\ref{lemma:opt_ratio}, there is an optimal policy $\phi^{*}_{s}\in\Phi_{s}$ that maximizes the problem defined in \eqref{eqn:sub_problem:1}, satisfying $\alpha^{\phi^{*}_s}_{j,s}(n) = 1$ for all $j\in\mathcal{J}$, $s\in\mathcal{S}_{j}$, $n\in\mathcal{C}_{k(s)}$. 
There is also a policy $\varphi_j\in\overline{\Phi}_{j}$, $j\in\mathcal{J}$, for the maximization problem defined by \eqref{eqn:sub_problem:2}, satisfying $\overline{a}^{\varphi_j}_{j}=1-A_{j}$ and $\alpha^{\varphi_j}_{j}=1-A_{j}$. 
Note that $1-A_{j} \in [0,1)$ under Condition~\ref{cond:heavy_traffic}.

Therefore, Constraints~\eqref{eqn:relax:constraint_1},
\eqref{eqn:relax:constraint_3} and \eqref{eqn:relax:constraint_4} are
satisfied with equality: the complementary slackness condition for the dual and primal problems is satisfied.
The optimal solution $\phi^{*}$ determined by $\phi^{*}_{s}$ ($s\in\mathcal{S}$) and $\varphi_j$ ($j\in\mathcal{J}$) that maximizes the Lagrangian problem defined in \eqref{eqn:dual} will also maximize the primal problem defined by \eqref{eqn:e_star:objective}, \eqref{eqn:relax:constraint_1}, \eqref{eqn:relax:constraint_3}-\eqref{eqn:relax:constraint_4}.

We now consider the case where $J=1$ and denote the only $j$ in $\mathcal{J}$ by $j^{*}$.
The proposition can be proved along the same lines as the $J>1$ case if $A_{j^{*}} \leq 1$.
If $A_{j^{*}}> 1$, then let
$\gamma_{j^{*}} = \max\left\{0,-\nu_{j^{*}}\right\}$.
From Lemma~\ref{lemma:opt_ratio}, there is an optimal policy
$\phi^{*}_{s}\in\Phi_{s}$ that maximizes the problem defined in
\eqref{eqn:sub_problem:1}, satisfying \eqref{eqn:prop:opt_equal:1} for
$j=j^{*}$, $s\in\mathcal{S}_{j^{*}}$ and
$n\in\mathcal{C}_{k(s)}$.  Since $A_{j^{*}}> 1$, there exists
a
$\nu_{j^{*}} \geq
\min_{s\in\mathcal{S}_{j^{*}}}\lambda_{j^{*}}(1-e^{*}(\varepsilon_{k(s)}-\varepsilon_{k(s)}^{0})/\mu_{k(s)})$,
such that
$
\sum_{s\in\mathcal{S}_{j^{*}}}\sum_{n\in\mathcal{B}_{k(s)}}\pi^{\phi^{*}_{s}}_{s}(n)\alpha^{\phi^{*}_s}_{s}(n)
=1$.
Note that $\phi^*_s$ is dependent on $\nu_{j^*}$.

 On the other hand, the setting of $\gamma_{j^{*}}$ guarantees
that there is  a policy $\varphi\in\overline{\Phi}_{j^{*}}$
maximizing the objective function defined by
\eqref{eqn:sub_problem:2}, and
satisfying $\alpha^{\varphi}_{j^{*}} = 0$ and
$\Theta(\overline{a}^{\varphi}_{j^{*}})=0=\Theta(1-A_{j^{*}})$,
e.g., equations \eqref{eqn:relax:constraint_1} and
\eqref{eqn:relax:constraint_3} achieve equality.  That is, such a
$\nu_{j^{*}}$, $\gamma_{j^{*}}$, $\phi^{*}_{s}\in\Phi_{s}$,
$s\in\mathcal{S}_{j^{*}}$ and
$\varphi\in\overline{\Phi}_{j^{*}}$, make the complementary
slackness condition satisfied.
\end{IEEEproof}

\vspace{-0.3cm}

\section{Proof of Proposition~\ref{prop:sup_equilibrium}}
\label{app:prop:sup_equilibrium}

Consider $J+|\mathcal{S}|$ sequences of positive reals $t^h_{s,m}$, $m=1,2,\ldots$, for $s=1,2,\ldots,J+|\mathcal{S}|$, where  $h\in\mathbb{N}_+$ is the scaling parameter. 
We define $t^h_{s,m}$ for $s=1,2,\ldots,J$ as time of the $m$th arrival of jobs of type $s$, and define $t^h_{s,m}$ for $s=J+1,J+2,\ldots,J+|\mathcal{S}|$ to be the time of the $m$th potential departure of jobs on server labeled by
$s-J$. Define $t^h_{s,0} = 0$ for any $s=1,2,\ldots,J+|\mathcal{S}|$.
For our network system, the inter-arrival and inter-departure times are positive with probability $1$ and, also with probability $1$,  no two events  occur at the same time.
Let $\tau(t)$ represent the latest event, either arrival or potential departure, that happens before time $t$.
We define a random vector $\bm{\xi}^{h}_{t}$, 
for $s=1,2,\ldots,|\mathcal{S}|+J$ and $t\geq 0$:
if $\tau(t^h_{s,m})\leq t < t^{h}_{s,m^*}$ where the $m^*$ satisfies $t^{h}_{s,m^*-1}\leq t < t^{h}_{s,m^*}$, then $\xi^{h}_{s,t}=1/(t^{h}_{s,m^*}-\tau(t^{h}_{s,m^*}))$;
otherwise, $\xi^{h}_{s,t} = 0$.
The sample paths of  $\bm{\xi}^{h}_t$ are almost surely continuous in $t\geq 0$, except for a finite number of discontinuities of the first kind in  a bounded period of $t>0$.
Let 
$\Lambda_{j,i}(\bm{x}) =u^{PAS,h}_{j,i}(\frac{\bm{x}}{hS_0})z_{i}hS_0$, which is independent from $h$ from the definition of $u^{PAS,h}_{j,i}(\cdot)$.
We define a function, $Q^{h}(i,i',\bm{x},\bm{\xi}^{h})$,  for
$h\in\mathbb{N}^{+}$, $n_i,n_{i'}\in\mathcal{B}_{k}$,
$k\in\mathcal{K}\cup\{0\}$, $\bm{x}\in\mathbb{R}^{I}$,
$\bm{\xi}^{h}\in\mathbb{R}^{hS_0+J}$, by:
if $k_{i}=k_{i'}$ and $i'=i+1$,\vspace{-0.2cm}
$$
Q^{h}(i,i',\bm{x},\bm{\xi}^{h})=\sum_{j\in\mathcal{J}_{k_{i}}}\left[\Lambda_{j,i}(\bm{x})\ 
+f^{0,h}_{i,a,j}(\bm{x})\right]\xi^{h}_{j};\vspace{-0.2cm}
$$
if $k_{i}=k_{i'}$ and $ i'=i-1$, \vspace{-0.2cm}
$$Q^{h}(i,i',\bm{x},\bm{\xi}^{h})=\sum\limits_{m=\lceil x^{-}_{i-1}\rceil+1}^{\lceil x^{-}_{i}\rceil}\xi^{h}_{m+J}
 + \sum\limits_{m=1}^{hS_0}f^{h}_{i,a,m}(\bm{x})\xi^{h}_{m+J};\vspace{-0.2cm}$$
otherwise, $Q^{h}(i,i',\bm{x},\bm{\xi}^{h})=0$;
where $\mathcal{J}_{k}= \{j\in\mathcal{J}|\ k\in\mathcal{K}_{j}\}$,
$k\in\mathcal{K}$, $x^{-}_{i}=\sum_{m=1}^{i}x_{m}$ with $x^{-}_{0}=0$,
and $\bm{f}^{0,h}_{i,a}(\bm{x})\in\mathbb{R}^{J}$ and
$\bm{f}^{h}_{i,a}(\bm{x})\in\mathbb{R}^{hS_0}$ with $0<a<1$ are appropriate functions as described in \cite{fu2018restless} such that
$Q^{h}$ is Lipschitz continuous
 in
$\bm{x}$. For the special case $h=1$ and for any given $0<a<1$,
$Q^{\phi,1}(i,i',\bm{x},\bm{\xi}^{1})$  satisfies a
Lipschitz condition over $\bm{x}\in \mathbb{R}^{I}$ and
$\bm{\xi}^{1}\in \mathbb{R}^{S_0+J}$.  For $0<a<1$ and
$\epsilon>0$, define 
\vspace{-0.3cm}
\begin{multline*}\label{eqn:differential_equation_x}
\dot{X}_{t,i}^{\epsilon}
\coloneqq\sum\limits_{i'=1}^{I}Q^{1}(i',i,\bm{X}^{\epsilon}_{t},\bm{\xi}_{t/\epsilon}^{1})-Q^{1}(i,i',\bm{X}^{\epsilon}_{t},\bm{\xi}_{t/\epsilon}^{1})\\
=(\overline{\bm{q}}_i(\bm{X}^{\epsilon}_{t}),\bm{\xi}^{1}_{t/\epsilon}),
\end{multline*}
where $\overline{\bm{q}}_i(\bm{x})$ is a vector of length $S_0+J$ dependent on $\bm{x}$ and $(\cdot,\cdot)$ represents inner product of vectors.
It follows that $\dot{\bm{X}}_{t}^{\epsilon}$ satisfies a Lipschitz condition over $\bm{X}^{\epsilon}_{t}$ and $\bm{\xi}^{1}_{t/\epsilon}$. 
Let 
\begin{equation*}
b(\bm{x}, \bm{\xi})\coloneqq ((\overline{\bm{q}}_i(\bm{x}),\bm{\xi}):i=1,2,\ldots,I)\coloneqq \overline{\mathcal{Q}}(\bm{x})\bm{\xi},
\end{equation*}
where where $\overline{Q}(\bm{x})$ is a $I\times (S_0+J)$ matrix. 
For any $\bm{x}\in \mathbb{R}^{I}$, $\delta>0$, there exists $\overline{b}(\bm{x})$ satisfying
\begin{equation}\label{eqn:equilibrium_b}
\lim\limits_{T\rightarrow +\infty} \mathbb{P}\biggl\{
\Bigl\lVert \frac{1}{T}\int_{t}^{t+T}b^{\phi}(\bm{x},\bm{\xi}^{1}_{s})ds-\overline{b}(\bm{x})\Bigr\rVert>\delta\biggr\} = 0,
\end{equation}
uniformly in $t>0$.
Let $\overline{\bm{x}}(t)$ be the solution of $\dot{\overline{\bm{x}}}(t)=\overline{b}(\overline{\bm{x}}(t))$ and $\overline{\bm{x}}(0)=\bm{X}^{\epsilon}_{0}=\bm{x}_0$.
Define, for $Y = S_0+J$, the random variables $\bm{\xi}^{1}_{t}$
and vectors $\bm{\alpha}_{t}\in\mathbb{R}^{Y}$, 
\begin{equation}\label{eqn:definition_H_xi}
\int_{0}^{T}H_{\xi}(\bm{\alpha}_{t})d\ t 
= \lim\limits_{\varepsilon \to 0}\varepsilon \ln \mathbb{E}\ \exp\Bigl\{\frac{1}{\varepsilon}\int_{0}^{T}(\bm{\alpha}_t,\bm{\xi}^{1}_{t/\varepsilon})dt \Bigr\}.
\end{equation}
We discuss the existence of $H_{\xi}(\cdot)$ satisfying \eqref{eqn:definition_H_xi} next.
Let  
$
\bm{\Upsilon}_{T} = \int_0^{T}\bm{\xi}^{1}_{t}dt,
$
which is a Poisson distributed random vector.
Writing
$\bm{\alpha}=(\alpha_1,\alpha_2,\ldots,\alpha_{Y})$, we obtain \vspace{-0.4cm}
\begin{equation}\label{eqn:H_xi}
H_{\xi}(\bm{\alpha})
=\sum_{j=1}^{J}\lambda_{j} (e^{\alpha_j}-1) +\sum_{s\in\mathcal{S}}\mu_{k(s)} (e^{\alpha_{J+s}}-1). 
\end{equation}

\vspace{-0.4cm}

It follows that $H_{\xi}(\bm{\alpha})$ satisfying \eqref{eqn:H_xi} is bounded for any bounded
$\bm{\alpha}$, so that the functional $H_{\xi}(\bm{\alpha}_{t})$ satisfying \eqref{eqn:definition_H_xi} also exists for any continuous $\bm{\alpha}_{t}$ on
$0\leq t\leq T$.
For a vector $\bm{U} \in(\mathbb{R}^{+})^{Y}$, define a function
$\mathcal{T}^{\bm{U}}$ for $i=1,2,\ldots,Y$ and $\bm{x}\in \mathbb{R}^{Y}$:
if $0\leq x_{i} \leq U_{i}$, $\mathcal{T}^{\bm{U}}_{i}(\bm{x})=x_{i}$; if $x_i > U_i$, $\mathcal{T}^{\bm{U}}_{i}(\bm{x})=U_{i}$; otherwise, $\mathcal{T}^{\bm{U}}_{i}(\bm{x})=0$.
For $\tilde{\bm{\alpha}},\bm{x}\in\mathbb{R}^{I}$, define\vspace{-0.2cm}
\begin{equation}\label{eqn:define_H}
H(\bm{x},\tilde{\bm{\alpha}})=
\lim\limits_{T \rightarrow +\infty}\frac{1}{T} \ln \mathbb{E} \exp\Bigl\{\int\limits_{0}^{T}(\tilde{\bm{\alpha}},\tilde{b}^{\bm{U}}(\bm{x},\bm{\xi}^{1}_{t}))dt\Bigr\} \vspace{-0.2cm}
\end{equation}
where 
$
\tilde{b}^{\bm{U}}(\bm{x},\bm{y})\coloneqq
b(\bm{x},\mathcal{T}^{\bm{U}}(\bm{y})) = \overline{\mathcal{Q}}(\bm{x})\mathcal{T}^{\bm{U}}(\bm{y}).
$
Here, since $b(\cdot,\cdot)$ is Lipschitz continuous in both
arguments, all of the elements of $\bm{U}$  are finite and positive, and for $\overline{\mathcal{Q}}(\cdot)=(q_{i,j})$, $|q_{i,j}|< +\infty$ for all $i=1,2,\ldots,I$ and $j=1,2,\ldots,Y$, 
we obtain that $\tilde{b}^{\bm{U}}(\cdot, \cdot) $ is bounded and Lipschitz continuous in both arguments.
From~\cite[Lemma 4.1, Chapter 7]{freidlin2012random}, $H(\bm{x},\bm{y})$ is jointly continuous in both arguments and convex in the second argument.

We obtain from \eqref{eqn:define_H}, for any $\tilde{\bm{\alpha}},\bm{x}\in\mathbb{R}^{I}$, 
$
|H(\bm{x},\tilde{\bm{\alpha}})| \leq 
H_{\xi}(\mathcal{A}(\tilde{\bm{\alpha}}\overline{\mathcal{Q}}(\bm{x}))) $,
where $\mathcal{A}(\bm{x})$ takes absolute values of all $\bm{x}$'s elements.
Recall that $\overline{\mathcal{Q}}(\bm{x})$,
$\bm{x}\in\mathbb{R}^{I}$, is Lipschitz continuous on $\bm{x}$. 
For any compact set $\mathcal{A}^c\subset \mathbb{R}^I$ and $\tilde{\bm{\alpha}}\in\mathcal{A}^c$, $H(\bm{x},\tilde{\bm{\alpha}})$ obtained from \eqref{eqn:define_H} is bounded and because of its joint continuity, is Riemann integrable. Hence, the $H(\bm{x},\tilde{\bm{\alpha}})$ defined in \eqref{eqn:define_H} satisfies \vspace{-0.2cm}
\begin{equation}
\int\limits_0^T H(\bm{x}_t,\tilde{\bm{\alpha}}_t)dt
=\lim\limits_{\varepsilon\to 0}\varepsilon \ln \mathbb{E} \text{exp}\Bigl\{\frac{1}{\varepsilon}\int\limits_0^T(\tilde{\bm{\alpha}}_t,\bm{\xi}^1_{t/\varepsilon})dt\Bigr\}. \vspace{-0.2cm}
\end{equation}

Now  we consider the Legendre transform of $H(\bm{x},\tilde{\bm{\alpha}})$:
\begin{equation}\label{eqn:legendre}
L(\bm{x},\bm{\beta}) = \sup_{\tilde{\bm{\alpha}}\in\mathbb{R}^{I}} \left[(\tilde{\bm{\alpha}},\bm{\beta})-H(\bm{x},\tilde{\bm{\alpha}})\right].
\end{equation}
$L(\bm{x},\bm{\beta})$ is strictly convex in the second argument
if $H(\bm{x},\tilde{\bm{\alpha}})$ is strictly convex in the
second argument. Let $\tilde{\bm{\alpha}}=\bm{0}$,
$(\tilde{\bm{\alpha}},\bm{\beta})-H(\bm{x},\tilde{\bm{\alpha}})
=0$, so that $L(\bm{x},\bm{\beta})$ is always non-negative.

\vspace{-0.2cm}
\begin{lemma}\label{lemma:average_vanish}
If $L(\bm{\varphi},\bm{\beta})$ is strictly convex in the second argument, then $L(\bm{\varphi},\bm{\beta}) = 0$ if and only if $\bm{\beta} = \mathbb{E} \tilde{b}^{\bm{U}}(\bm{\varphi},\bm{\xi}^{1}_{t})$.
\end{lemma}  \vspace{-0.2cm}
\begin{IEEEproof}
The $L(\bm{\varphi},\bm{\beta})=0$ when $\bm{\beta} = \mathbb{E}
\tilde{b}^{\bm{U}}(\bm{\varphi},\bm{\xi}^{1}_{t})$,~\cite[Chapter 7, Section 4]{freidlin2012random}. 
Together with non-negativity and strict convexity of $L(\bm{\varphi},\bm{\beta})$, for a given $\bm{\varphi}\in\mathbb{R}^{I}$,  $L(\bm{\varphi},\bm{\beta})=0$ if and only if  $\bm{\beta}=\mathbb{E} \tilde{b}^{\bm{U}}(\bm{\varphi},\bm{\xi}^{1}_{t})$.
\end{IEEEproof}
The second derivative $\partial^{2}H/\partial\tilde{\bm{\alpha}}^2$ exists and is continuous in $\bm{U}$, and 
as $\bm{U}\to\bm{ \infty}$, 
it converges point-wisely to $ \partial^2 H_{\xi}(\tilde{\bm{\alpha}}\overline{Q}(\bm{x}))/\partial \tilde{\bm{\alpha}}^2$; the function $H_{\xi}(\tilde{\bm{\alpha}}\overline{Q}(\bm{x}))$ is
strictly convex in $\tilde{\bm{\alpha}}$ by
\eqref{eqn:H_xi} and the second derivative in $\tilde{\bm{\alpha}}$ exists.
Thus, for sufficiently large $\bm{U}$, $L(\bm{\varphi},\bm{\beta})$ is always strictly convex in the second argument. 
\begin{IEEEproof}
Let
$S_{0,T}(\varphi)=\int_{0}^{T}L(\bm{\varphi}_t,\dot{\bm{\varphi}}_{t})d\
t$, where $\varphi$ denotes a trajectory
$\bm{\varphi}_{t}\in\mathbb{R}^{I}$ ($0\leq t\leq T$), and let $C_{0,T}$ represent the compact set of all such trajectories with $\bm{\varphi}_0=\bm{x}_0\in\mathbb{R}^I$. 
Define a closed set $A(\bm{U},\delta)\coloneqq \{\varphi\in C_{0,T}|\lVert\varphi_t-\tilde{\bm{x}}^{\bm{U}}_t\rVert \geq \delta\}$ where $\tilde{\bm{x}}^{\bm{U}}_t$ is the solution of
$\dot{\tilde{\bm{x}}}^{\bm{U}}_t = \mathbb{E}
\tilde{b}^{\bm{U}}(\bm{\varphi},\bm{\xi}^{1}_{t})$ with $
\tilde{\bm{x}}^{\bm{U}}_0 = \bm{x}_0$.
From~\cite[Theorem 4.1 in Chapter 7 \& Theorem 3.3 in Chapter 3]{freidlin2012random}, 
for any $\bm{U}\in(\mathbb{R}^+)^Y$ and $\delta >0$, \begin{equation*}
\overline{\lim\limits_{\epsilon \to 0}}~\epsilon\ln \mathbb{P}\Bigl\{\sup\limits_{0\leq t\leq T} \bigl\lVert \tilde{\bm{X}}^{\bm{U},\epsilon}_{t}-\tilde{\bm{x}}^{\bm{U}}_t\bigr\rVert >\delta\Bigr \}
\leq 
-\inf_{\varphi\in A(\bm{U},\delta)}S_{0,T}(\varphi),\end{equation*}
where $\tilde{\bm{X}}^{\bm{U},\epsilon}_{t}$ is the solution of  \vspace{-0.2cm}
\begin{equation*}
\dot{\tilde{\bm{X}}}_{t}^{\bm{U},\epsilon}=\tilde{b}^{\bm{U}}(\tilde{\bm{X}}_{t}^{\bm{U},\epsilon},\bm{\xi}^{1}_{t})),~\tilde{\bm{X}}_{0}^{\bm{U},\epsilon}=\bm{x}_0. \vspace{-0.3cm}
\end{equation*}
Also, because \vspace{-0.2cm}
\begin{equation*}
\lim_{\bm{U}\rightarrow \bm{\infty}}\tilde{\bm{x}}^{\bm{U}}_t = \overline{\bm{x}}(t)~\text{and}~\lim_{\bm{U}\rightarrow \bm{\infty}}\tilde{\bm{X}}^{\bm{U},\epsilon}_t =\bm{X}^{\epsilon}_t,\vspace{-0.2cm}
\end{equation*}
by Lemma~\ref{lemma:average_vanish}, we obtain, for any $\delta >0 $, 
there exist $s>0$ and $\epsilon_0>0$ such  that, for  all positive $\epsilon<\epsilon_0$, 
\vspace{-0.2cm}
\begin{equation}\label{eqn:sup_equilibrium_1}
\mathbb{P}\Bigl\{ \sup\limits_{0\leq t\leq T} \lVert\bm{X}^{\epsilon}_{t}-\overline{\bm{x}}(t)\rVert>\delta\Bigr\}  \leq 
e^{-\frac{s}{\epsilon}}.\vspace{-0.2cm}
\end{equation}

Recall that functions $Q^h$ and $b$ are depenedent on a parameter $a\in(0,1)$.
Equation \eqref{eqn:sup_equilibrium_1} holds for any given
$a$.  
Because of the Lipschitz behavior of
$\dot{\bm{X}}^{\epsilon}_{t}$ and $\dot{\overline{\bm{x}}}(t)$ on
$0<a<1$, $\lim_{a\to 0}d \dot{\bm{X}}^{\epsilon}_{t}/d a = 0$ and
$\lim_{a\to 0}d \dot{\overline{\bm{x}}}(t)/d a = 0$,
Equation~\eqref{eqn:sup_equilibrium_1} also holds in the limiting case $a\to 0$. By slightly abusing notation, in the following, we still use $Q^h$, $b$ and $\bar{\bm{x}}(t)$ to represent $\lim_{a\to 0}Q^h$, $\lim_{a\to 0}b$ and $\lim_{a\to 0}\bar{\bm{x}}(t)$.

Along similar lines to~\cite{fu2016asymptotic}, we interpret the scalar $\epsilon$ and the scaling effects in another way. 
For $\bm{x}\in \mathbb{R}^{I}$ and $\bm{\xi}^{h}\in \mathbb{R}^{hS_0+J}$, define \vspace{-0.2cm}
\begin{equation*}\label{eqn:differential_equation_xn}
b^{h}(\bm{x}, \bm{\xi}^{h})\coloneqq\sum_{i'=1}^{I}Q^{h}(i',i,\bm{x},\bm{\xi}^{h})-Q^{h}(i,i',\bm{x},\bm{\xi}^{h}). \vspace{-0.2cm}
\end{equation*} 
If we set $\epsilon = 1/h$, then following the same technique
as~\cite{fu2016asymptotic}, for any $\bm{x}\in \mathbb{R}^{I}$,
$h\in\mathbb{N}^{+}$ and $T>0$, we observe that
$\int_{0}^{T}b(\bm{x},\xi^{1}_{t/\epsilon})d\ t$ and
$\int_{0}^{T}(b^{h}(h\bm{x},\xi^{h}_{t})/h)d\ t$ are
identically distributed.
Define $\bm{Z}^{\epsilon}_0= \bm{Z}^{h}_0=\bm{x}_{0}/S_0=\bm{z}^0$, and
\begin{equation*}
\dot{\bm{Z}}^{h}_t = \frac{b^{h}(hS_0\bm{Z}^{h}_t,\bm{\xi}^{h}_{t})}{hS_0}
\text{ and }
\dot{\bm{Z}}^{\epsilon}_{t} = \frac{b(S_0\bm{Z}^{\epsilon}_{t},\bm{\xi}^{1}_{t/\epsilon})}{S_0}.
\end{equation*}
From \eqref{eqn:sup_equilibrium_1}, for any $T>0$, $\delta >0 $, 
there exist positive  $s$ and $H$ such  that, for  all $h>H$, 
\begin{equation}\label{eqn:sup_equilibrium_2}
\mathbb{P}\biggl\{ \sup\limits_{0\leq t\leq T} \Bigl\lVert\bm{Z}^{h}_t-\overline{\bm{x}}(t)/S_0\Bigr\rVert>\delta\biggr\} \leq e^{-sh}.
\end{equation}
Effectively then, scaling time by $\epsilon = \frac{1}{h}$ is equivalent to scaling system size by $h$.
From
\eqref{eqn:sup_equilibrium_2}, Corollary~\ref{coro:equilibrium_vector} and \cite[Lemma 3]{fu2018restless}, for any $T>0$, $\delta >0 $, there
exist $s>0$ and $H>0$ such that for any $h > H$, \eqref{eqn:sup_equilibrium_3} holds.
\end{IEEEproof}
\vspace{-0.5cm}
\section{Settings for Simulations in Figure~\ref{fig:fig2}}\label{app:fig2}
\vspace{-0.2cm}
Define Case~I as a system with same settings as in the simulations
  for Figure~\ref{fig:fig1a}, except for the following parameters:
\begin{itemize}
\item $\mu_1=8.06114$, $\varepsilon_1^0=0.80611$, $\varepsilon_1=8.06114$;
\item $\mu_2=4.05127$, $\varepsilon_2^0=0.94774$,   $\varepsilon_2=4.73868$;
\item $\mu_3=3.70788$, $\varepsilon_3^0=2.21086$, $\varepsilon_3=7.36952$;
\item $\mu_4=2.88018$, $\varepsilon_4^=4.40851$, $\varepsilon_4=11.02129$;
\item $\mu_5=2.44950$, $\varepsilon_5^0=8.79314$, $\varepsilon_5=17.58627$;
\item and $\lambda_1=6.30639$, $\mathcal{K}_1=\{1,5\}$.
\end{itemize}

Define Case~II as a system with same settings as in the simulations
  for Figure~\ref{fig:fig1b}, except for the following parameters:
\begin{itemize}
\item $\mu_1=7.74574$, $\varepsilon_1^0=0.77457$, $\varepsilon_1=7.74574$;
\item $\mu_2=7.46818$, $\varepsilon_2^0=1.77913$, $\varepsilon_2=8.89565$;
\item $\mu_3=6.32019$, $\varepsilon_3^0=2.65737$, $\varepsilon_3=8.85791$;
\item $\mu_4=4.66817$, $\varepsilon_4^0=3.27786$, $\varepsilon_4=8.19465$;
\item $\mu_5=4.62779$, $\varepsilon_5^0=4.82040$, $\varepsilon_5=9.64079$;
\item $\lambda_1=11.04970$, $\mathcal{K}_1=\{2,3,5\}$;
\item $\lambda_2=8.27302$, $\mathcal{K}_2=\{2,3\}$;
\item $\lambda_3=11.04970$, $\mathcal{K}_3=\{2,3,5\}$.
\end{itemize}
Note that both examples are taking instances of the randomly generated system in simulations in Sections~\ref{subsec:one_job} and \ref{subsec:multi_job}.
\vspace{-0.5cm}
\section{Settings for Simulations in Section~\ref{subsec:google}}\label{app:fig_google}
\vspace{-0.2cm}
Consider ten server groups with the following parameters:
\begin{itemize}
\item $\mu_1=2.42523$, $\varepsilon_1^0=0.06548$, $\varepsilon_1=0.242523$;
\item $\mu_2=2.41588$, $\varepsilon_2^0=0.04974$, $\varepsilon_2=0.207254$;
\item $\mu_3=2.38966$, $\varepsilon_3^0=0.04279$, $\varepsilon_3=0.20377$;
\item $\mu_4=2.22434$, $\varepsilon_4^0=0.02630$, $\varepsilon_4=0.14613$;
\item $\mu_5=1.75822$, $\varepsilon_5^0=0.07228$, $\varepsilon_5=0.60241$;
\item $\mathcal{K}_1=\{1,5,6,10\}$, $B_1 = 20$;
\item $\mathcal{K}_2=\{1,2,3,4,5,7,8,9\}$, $B_2 = 20$; 
\item $\mathcal{K}_3=\{1,6,7,10\}$, $B_3 = 20$; 
and $\mathcal{K}_4 =\{2\}$, $B_4 = 20$.
\end{itemize} 
Note that all the numbers are generated by a pseudo-random number generator, and the unit of service rates $\mu_k$ is $10^{-4}s^{-1}$: the number of processed jobs per second. The values of $\mu_k$ are normalized to be sufficiently small that we can observe a positive number of blocked jobs in Figure~\ref{fig:google-blocking}, and the heavy traffic condition can be achieved during the peak hours as presented in Section~\ref{subsec:google}.

\vspace{-0.3cm}
\section*{Acknowledgment}

Jing Fu's research is supported by the Australian Research Council (ARC) Centre of Excellence for the Mathematical and Statistical Frontiers (ACEMS) and ARC Laureate Fellowship FL130100039.
\vspace{-0.4cm}

\ifCLASSOPTIONcaptionsoff
  \newpage
\fi

\bibliographystyle{references/IEEEtran}
%\bibliography{references/IEEEabrv,references/bib1}

\end{document}